\documentclass[aps,prc,twocolumn,nofootinbib,amsmath,amssymb,10pt]{revtex4-2}
\usepackage[utf8]{inputenc}
\usepackage{amsmath}
\usepackage{graphicx}
\usepackage{float}
\usepackage{hyperref}
\usepackage{lipsum}
\usepackage{listings}
\usepackage{listingsutf8}
\usepackage{xcolor}
\usepackage{amssymb}
\usepackage{textcomp}
\usepackage{units}
\usepackage{bbm} 
\usepackage{url}
\usepackage[english]{isodate}
\usepackage{bm}
\usepackage{mdframed}
\usepackage[capitalise]{cleveref}
\usepackage{enumerate}
\usepackage{dsfont}
\usepackage{dcolumn}
\usepackage{physics}
\usepackage{color}
\usepackage{placeins}
\usepackage{slashed}
\usepackage{amssymb}
\usepackage{hyperref}
\usepackage[capitalise]{cleveref}
\usepackage[caption=false]{subfig}
\usepackage{braket}

\renewcommand{\l}{\ell}
\newcommand{\NN}{NN}

\newcommand*{\tS}{^3S_1 \mathrm{-}^3D_1}
\newcommand*{\tP}{^3P_2 \mathrm{-}^3F_2}

\newcommand*{\LO}{LO}
\newcommand*{\NLO}{NLO}
\newcommand*{\NNLO}{N$^2$LO}
\newcommand*{\NNNLO}{N$^3$LO}

\newcommand*{\VLO}{V^{(0)}}

\newcommand*{\TLO}{T^{(0)}}

\newcommand*{\TNNLO}{T^{(2)}}

\newcommand*{\Tl}{T_\mathrm{lab}}
\newcommand*{\CM}{c.m.\@}

\newcommand*{\exlam}{\bar{\Lambda}}

\newcommand*{\pon}{k}

\newcommand{\mN}{m_N}
\newcommand{\cm}{c.m.}
\newcommand{\Nmax}{N_\mathrm{max}}

\newcommand{\N}{\mathcal{N}}
\newcommand{\J}{\mathcal{J}}
\renewcommand{\L}{\mathcal{L}}

\newcommand{\chEFT}{$\chi$EFT}

\usepackage{todonotes}

\begin{document}

\title{Perturbative $\chi$EFT calculations of the deuteron and triton up to N$^2$LO}

\author{Oliver Thim} \email{oliver.thim@chalmers.se}
\author{Andreas Ekström}
\author{Christian Forssén}
\affiliation{Department of Physics, Chalmers University of Technology, SE-412 96, Göteborg, Sweden}

\date{\today}
\noindent
\begin{abstract} 
We extend previous studies of the deuteron and triton ground-state energies to next-to-next-to-leading order (\NNLO) in chiral effective field theory, employing a power counting in which subleading interactions are treated perturbatively. Triton calculations are performed using the no-core shell model, and we demonstrate converged perturbative results for regulator cutoffs up to $\Lambda \approx 1200$~MeV.
We analyze exceptional cutoffs in the $^3P_0$ and $\tS$ nucleon-nucleon channels and find a resulting cutoff dependence in the triton ground-state energy at \NNLO. The effect associated with the exceptional cutoff in the $^3P_0$ channel can be mitigated by redefining the leading-order wave function within the freedom allowed by the effective field theory. 
The same approach applied in the $\tS$ channel remedies the effect of this exceptional cutoff on the ground state energy of the deuteron, but not for the triton.
\end{abstract}

\maketitle

\newpage

\newpage
\section{Introduction}

Chiral effective field theory (\chEFT) promises a systematically improvable framework for deriving nuclear interaction potentials consistent with quantum chromodynamics \cite{Weinberg:1978kz,Weinberg:1990rz,Weinberg:1991um}, and in particular the spontaneously broken chiral symmetry \cite{Machleidt:2011zz,Epelbaum:2008ga,Hammer:2019poc}. A power counting (PC) scheme facilitates the organization of nucleon-interaction Feynman diagrams in order of decreasing importance. Each diagram carries a scaling $(Q/\Lambda_b)^\nu$, where $\nu$ is the chiral order, $Q$ represents a typical low-energy scale, e.g., the external nucleon momenta, and $\Lambda_b$ is the \chEFT{} breakdown scale, typically estimated to be 
$\Lambda_b \approx 500$--$600$~MeV~\cite{Melendez:2019izc,Millican:2024yuz} or possibly even lower~\cite{Thim:2024yks}.

In Weinberg PC (WPC) \cite{Weinberg:1990rz,Weinberg:1991um}, the potential is constructed from the sum of all relevant irreducible nucleon-contact and finite-range interactions up to a given chiral order. This potential is then iterated non-perturbatively using the Lippmann-Schwinger or Schrödinger equation, where a cutoff on the order of the breakdown scale is applied to regulate the appearing divergences. This is the predominant approach used in \textit{ab inito} \cite{Ekstrom:2022yea,Hergert:2020bxy} computations of nuclear observables. Quantitative potentials have been developed up to the fifth chiral order within this scheme \cite{Entem:2015xwa,Reinert:2017usi}.

Amplitudes and observables computed in WPC are sensitive to the value of the employed cutoff \cite{Nogga:2005hy}, which indicates an unphysical regulator dependence of the results. Removing this, such that only a residual higher-order dependence remains, is referred to as attaining renormalization group (RG) invariance \cite{vanKolck:2020llt}. In the nucleon-nucleon (\NN) sector, this can be achieved by promoting nucleon-contact interactions to lower chiral orders together with treating subleading interactions perturbatively \cite{Long:2007vp}. 

There are several ongoing efforts to construct PCs that render observables cutoff independent~\cite{PavonValderrama:2005uj,PavonValderrama:2005gu,PavonValderrama:2011fcz,Long:2013cya,PavonValderrama:2016lqn,Valderrama:2009ei,Birse:2005um,Long:2012ve,SanchezSanchez:2017tws,PhysRevC.85.034002,Yang:2016brl,Mishra:2021luw,Peng:2021pvo}. In particular, it has been shown that the PC proposed by Long and Yang \cite{Long:2012ve,PhysRevC.85.034002,Long:2011qx,Long:2007vp} performs well in the \NN{} sector up to $\nu=3$ (next-to-next-to-next-to-leading order \NNNLO) \cite{Thim:2023fnl,Thim:2024yks,Thim:2024jdv}. Furthermore, \citet{Song:2016ale} studied the triton up to next-to-leading order (NLO) and \citet{Yang:2020pgi} presented the first studies of $^4$He, $^6$Li, and $^{16}$O, also up to NLO. In the latter study, issues of $\alpha$-cluster instability were encountered,  possibly remedied by promoting few-nucleon forces to leading order (\LO)~\cite{Yang:2021vxa}. 

The cutoff dependence in the Long and Yang PC was examined in detail by \citet{Gasparyan:2022isg} who found that the low-energy constants (LECs) for the next-to-next-to-leading order (N$^2$LO) contact interaction in the $^3P_0$ channel diverge at certain \emph{exceptional} values of the cutoff. This effect is related to the perturbative calculation of amplitudes, and appears to break cutoff independence beyond next-to-leading order (\NLO{}) since the divergences in the LECs propagate to predicted observables. Exceptional cutoffs were further studied in Refs.~\cite{Peng:2024aiz,Yang:2024yqv}, where it was demonstrated that cutoff independence can be restored for \NN{} scattering amplitudes by utilizing the EFT freedom to adjust the LO wave function from which the subleading amplitudes are perturbatively computed. However, the effect of exceptional cutoffs beyond the \NN{} system is unknown, as they first appear at \NNLO, for which no calculations have been performed. 

In this work, we extend perturbative \chEFT{} calculations for the triton, using the no-core shell model (NCSM)~\cite{Barrett:2013nh}, up to \NNLO{} in the Long and Yang PC. We investigate the approach of modifying the LO wave function to mitigate the impact of exceptional cutoffs in the triton ground-state energy. To this end, we also extend the analysis of exceptional cutoffs to the coupled $\tS$ channel, see also \cite{Peng:2025ykg}.

The article is organized as follows. In Sec.~\ref{sec:interactions} we briefly discuss the Long and Yang PC up to \NNLO{} as well as our calibration of the LECs. The theory of exceptional cutoffs, as presented in Refs.~\cite{Gasparyan:2022isg,Peng:2024aiz,Yang:2024yqv}, is summarized in Sec.~\ref{sec:exc-nn} and extended to the coupled $\tS$ channel, with application to the deuteron ground-state energy. In Sec.~\ref{sec:triton} we present perturbative computations of the triton ground-state energy up to \NNLO{}. Finally, Sec.~\ref{sec:summary} contains the conclusions and outlook.

\section{The nuclear interaction up to \NNLO{} in the Long and Yang PC\label{sec:interactions}}

Here, we briefly summarize the construction of the potentials in the Long and Yang  PC~\cite{Long:2012ve,PhysRevC.85.034002,Long:2011qx,Long:2007vp} and our calibration of the LEC values. We will follow the procedure of Ref.~\cite{Thim:2024yks}, which we refer to for further details.

The relevant pion- and contact-potential contributions up to \NNLO{} are summarized in \Cref{tab:potentials_PC}. The potential at \LO{} is treated nonperturbatively and consists of one-pion exchange (OPE), $V^{(0)}_{1\pi}$, together with contacts, $V^{(0)}_\mathrm{ct}$, parametrized by the LECs: $C^{(0)}_{^1S_0},\,C^{(0)}_{^3S_1},\,D^{(0)}_{^3P_0},\,D^{(0)}_{^3P_2}$.
In channels where OPE is singular and attractive, nonperturbative iteration requires counterterms with associated LECs to ensure cutoff independence~\cite{Nogga:2005hy}. This is why some $P$-wave counterterms have been promoted to \LO{}, where WPC only prescribes counterterms in $S$-waves. The short-distance singularity of OPE becomes increasingly suppressed for higher orbital angular momentum, which eventually enables a purely perturbative treatment~\cite{Birse:2005um,PhysRevC.99.024003,Peng:2020nyz}. As a result, only a limited set of channels must be iterated nonperturbatively at \LO. These channels are listed in the leftmost column of \Cref{tab:LECs}. 

\Cref{tab:potentials_PC} further shows the \NLO{} potential, $V^{(1)}$, which consists of two contact interactions in $^1S_0$ $(V^{(1)}_\mathrm{ct})$ together with OPE as the first contribution in the purely perturbative NN channels. At \NNLO{}, the leading two-pion exchange $(V^{(2)}_{2\pi})$ enters with associated contacts $(V^{(2)}_\mathrm{ct})$ but there is no contribution in the purely perturbative channels. \Cref{tab:LECs} lists all the LECs that parameterize the contact interactions from \LO{} to \NNLO{}.

All potentials are expressed in momentum space, and we employ a non-local regulator 
\begin{equation}
    V^{(\nu)}(\bm{p}',\bm{p}) \to e^{-p'^6/\Lambda^6} \ V^{(\nu)}(\bm{p}',\bm{p}) e^{-p^6/\Lambda^6},
    \label{eq:inc_reg}
\end{equation}
where $\Lambda$ is the momentum cutoff. Here, $\bm{p}$ ($\bm{p}'$) denotes the ingoing (outgoing) relative \NN{} momentum in the center-of-mass (\cm) frame and $p=|\bm{p}|$.

We calibrate the values of the LECs using \NN{} phase shifts computed in distorted-wave perturbation theory and data from the Nijmegen partial-wave analysis~\cite{Stoks:1993tb}, see Ref.~\cite{Thim:2024yks}.

\begin{table}
    \caption{Potential contributions in NN channels where OPE is treated non-perturbatively and perturbatively, respectively. Detailed expressions, numerical values for constants, and more discussion about the potential structure can be found in Ref.~\cite{Thim:2024yks}.}
\centering
\begingroup
\renewcommand{\arraystretch}{1.35} 
\begin{tabular}{l|c|c|c}
    & & non-perturbative (at LO) & purely perturbative \\
    order & potential & channels & channels \\
    \toprule
    LO & $V^{(0)}$ & $V^{(0)}_{1\pi} + V^{(0)}_{\mathrm{ct}}$ & 0 \\
    NLO & $V^{(1)}$ & $V^{(1)}_{\mathrm{ct}}$ & $V^{(0)}_{1\pi}$ \\
    \NNLO{} & $V^{(2)}$ & $V^{(2)}_{2\pi} + V^{(2)}_{\mathrm{ct}}$ & 0 \\
    \hline
\end{tabular}
\endgroup
\label{tab:potentials_PC}
\end{table}

\begin{table}
    \caption{LECs present in \NN{} channels up to \NNLO{}, see Ref.~\cite{Thim:2024yks} for details.}
\centering
\begingroup
\renewcommand{\arraystretch}{1.6} 
\begin{tabular}{c|c|c|c}
    Channel & LO & NLO & N$^2$LO \\
    \hline\hline
    $^1S_0$ & $C^{(0)}_{^1S_0}$ & $C^{(1)}_{^1S_0}$, $D^{(0)}_{^1S_0}$ & $C^{(2)}_{^1S_0}$, $D^{(1)}_{^1S_0}$, $E^{(0)}_{^1S_0}$ \\ 
     $^3P_0$ &$D^{(0)}_{^3P_0}$ & -& $D^{(1)}_{^3P_0}$,$E^{(0)}_{^3P_0}$  \\
     $^1P_1$ &- & -& $D^{(0)}_{^1P_1}$  \\
    $^3P_1$ &- & -& $D^{(0)}_{^3P_1}$  \\
    $\tS$ &$C^{(0)}_{^3S_1}$ &-&$C^{(1)}_{^3S_1}$,$D^{(0)}_{^3S_1}$,$D^{(0)}_{SD}$\\ 
   
    $\tP$ &$D^{(0)}_{^3P_2}$ & -& $D^{(1)}_{^3P_2}$,$E^{(0)}_{^3P_2},E^{(0)}_{PF}$  \\
    \hline
\end{tabular}
\endgroup
\label{tab:LECs}
\end{table}

\section{Exceptional cutoffs in the two-nucleon system\label{sec:exc-nn}}

In this section, we examine the emergence of exceptional cutoff values, denoted $\exlam$, at which subleading LECs diverge in a manner that also leads to divergences in predictions for \NN{} observables. The appearance of exceptional cutoffs was first studied by~\citet{Gasparyan:2022isg} who traced this effect to the oscillatory nature of the \LO{} wave function at $r\rightarrow 0$, where we note that $r \lesssim \Lambda^{-1}_b$ is outside the domain of applicability of the EFT expansion. Exceptional cutoffs were further investigated in Refs.~\cite{Peng:2024aiz,Yang:2024yqv}, where two similar approaches were proposed to remedy these divergences within an EFT framework. Below, we summarize the origin of exceptional cutoffs, using $^3P_0$ as a first example, and extend the analysis to the $\tS$ channel, see also Ref.~\cite{Peng:2025ykg}.

\subsection{The $^3P_0$ channel\label{sec:3p0}}

The \LO{} scattering amplitude in the $^3P_0$ channel, $\TLO$, is computed non-perturbatively by solving the Lippmann-Schwinger (LS) equation
\begin{equation}
    \TLO = \VLO + \VLO G^+_0\TLO,
    \label{eq:LS_op}
\end{equation}
with the free resolvent $G^+_0 = \left(E - H_0 + i\epsilon\right)^{-1}$, $H_0 = \bm{p}^2/m_N$, and the nucleon mass  $m_N= 2m_p m_n /(m_p + m_n)$ is defined in terms of the proton and neutron masses. The explicit expression of the LS equation in the $^3P_0$ channel reads
\begin{align}
    &\TLO(p',p;k) = \VLO(p',p) \nonumber\\ &+\int_0^\infty dq \ q^2 \ \VLO(p,q) \frac{m_N}{k^2-q^2+i\epsilon} \TLO(q,p;k),
\end{align}
where $k$ is the on-shell momentum and the \LO{} potential $\VLO$ is projected to the $^3P_0$ channel.

The NLO correction in the $^3P_0$ channel vanishes while the \NNLO{} contribution to the scattering amplitude reads
\begin{equation}
    \TNNLO = \Omega^\dagger_- V^{(2)}_{2\pi}\Omega_+ + \Omega^\dagger_-V^{(2)}_\mathrm{ct} \Omega_+  \equiv T^{(2)}_\pi + T^{(2)}_\mathrm{ct}.
    \label{eq:TNNLO_decomp}
\end{equation}
The parts containing the two-pion exchange and the contact terms can be separated by linearity, and the M\o ller wave operators are defined as $\Omega_+ = \mathds{1} + G^+_0\TLO$ and $\Omega^\dagger_- = \mathds{1} + \TLO G^+_0$. Following Ref.~\cite{Gasparyan:2022isg}, the contact part of the on-shell \NNLO{} amplitude in the $^3P_0$ channel can be decomposed as
\begin{equation}   
   T^{(2)}_{\text{ct}}(\pon,\pon;\pon) = D^{(1)}_{^3P_0}(\Lambda) \ \psi^2_\Lambda(\pon) + E^{(0)}_{^3P_0}(\Lambda) \ 2\psi_\Lambda(\pon)\psi'_\Lambda(\pon) 
   \label{eq:Tct_N2LO_3P0}
\end{equation}
where $\psi_\Lambda(\pon)$ and $\psi'_\Lambda(\pon)$ are closely related to the LO radial wave function, $R^{(0)}(kr)$, at the origin. For sharp cutoffs, as employed in \cite{Peng:2024aiz}, the relations read
\begin{equation}
    \frac{dR^{(0)}(kr)}{dr} \Big{|}_{r=0} \propto \psi_\Lambda(k) , \quad 
    \frac{d^3 R^{(0)}(kr)}{dr^3}\Big{|}_{r=0} \propto \psi'_\Lambda(k).
\end{equation}

The \NNLO{} contribution to the $^3P_0$ phase shift is computed as \cite{Thim:2024yks}
\begin{equation}
    \delta^{(2)}(\pon) = - \frac{\rho(\pon) T^{(2)}(\pon,\pon;\pon)}{2i\exp\left(2i\delta^{(0)}(\pon)\right)},
    \label{eq:delta2_uncoup}
\end{equation}
where $\delta^{(0)}$ is the LO $^3P_0$ phase shift and $\rho(k) = i\pi m_N k$. Since $\delta^{(2)}$ is linear in $T^{(2)}$ we can define the contribution $\delta^{(2)}_\pi$ ($\delta^{(2)}_\mathrm{ct}$) from $T^{(2)}_\pi$ ($T^{(2)}_\mathrm{ct}$) such that $\delta^{(2)} = \delta^{(2)}_\pi + \delta^{(2)}_\mathrm{ct}$.

The renormalization conditions to fix the \NNLO{} LECs, $D^{(1)}_{^3P_0}$ and $E^{(0)}_{^3P_0}$, reads \cite{Thim:2024yks}
\begin{align}
    \delta^{(2)}(k_1) &= 0 \\
    \delta^{(0)}(k_2) + \delta^{(2)}(k_2) &= \delta_\mathrm{exp}(k_2).
\end{align}
The on-shell momenta $k_1$ and $k_2$ correspond to laboratory kinetic energies $\Tl=25$~MeV and $\Tl=50$~MeV, respectively, and $\delta_\mathrm{exp}$ is the Nijmegen phase shift \cite{Stoks:1993tb}. The conditions can be formulated as a linear system in the LECs
\begin{align}
    &\begin{pmatrix}
        \bar{\psi}^2_\Lambda(k_1) & 2\bar{\psi}_\Lambda(k_1)\bar{\psi}'_\Lambda(k_1) \\ \bar{\psi}^2_\Lambda(k_2) & 2\bar{\psi}_\Lambda(k_2)\bar{\psi}'_\Lambda(k_2)
    \end{pmatrix} \begin{pmatrix}
        D^{(1)}_{^3P_0} \\ E^{(0)}_{^3P_0}
    \end{pmatrix} \nonumber \\ &= \begin{pmatrix}
        \delta^{(2)}_\pi(k_1)/\rho(k_1) \\ \left(\delta^{(0)}(k_2)+\delta^{(2)}_\pi(k_2)-\delta_\mathrm{exp}(k_2) \right)/\rho(k_2)
    \end{pmatrix}.
    \label{eq:A_3P0_in_text}
\end{align}
where we define two real-valued quantities $\bar{\psi}_\Lambda(k) \equiv \psi_\Lambda(k) e^{-i\delta^{(0)}(k)}$ and $\bar{\psi}'_\Lambda(k) \equiv \psi'_\Lambda(k) e^{-i\delta^{(0)}(k)}$ \cite{Gasparyan:2022isg}.

\begin{figure}
    \centering
    \includegraphics[width=\columnwidth]{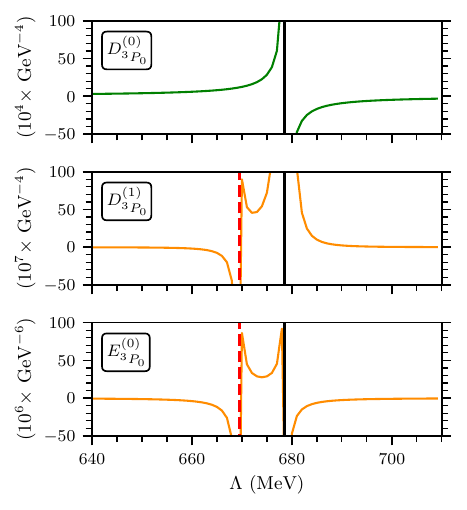}
    \caption{Values of the LECs at LO (top panel) and \NNLO{} (remaining panels) in the $^3P_0$ channel as a function of the momentum cutoff, $\Lambda$. The solid vertical line marks the location of the limit-cycle-like cutoff, while the dashed vertical line marks the exceptional cutoff in the given cutoff interval.}
    \label{fig:LECs_3P0}
\end{figure}

The determinant of the matrix in \cref{eq:A_3P0_in_text} can approach zero as the cutoff is varied, which leads to a singular system. There are two scenarios in which this can happen: 
\begin{enumerate}
    \item The function $\bar{\psi}_\Lambda(k) \to 0$, for all $k$. This occurs when the LO LEC, $D^{(0)}_{^3P_0}$, exhibits a limit-cycle–like divergence. For the \LO{} potential in this work a first limit-cycle-like divergence appears at $\Lambda=679$~MeV, as shown in the top panel of \cref{fig:LECs_3P0}. This type of zero is not problematic since the product of the diverging LEC and the vanishing short-distance piece of the wave function remains finite, rendering observables cutoff independent. Spurious deeply bound NN states are appearing at limit-cycle-like divergences \cite{Nogga:2005hy}. However, these can be projected out before carrying out calculations in the triton, see Sec.~\ref{sec:triton}.
    
    \item The determinant can have a non-trivial zero when
    \begin{equation}
        \frac{\bar{\psi}_\Lambda(k_1)\bar{\psi}'_\Lambda(k_2) - \bar{\psi}_\Lambda(k_2)\bar{\psi}'_\Lambda(k_1)}{\bar{\psi}_\Lambda(k_1)} = 0.
    \end{equation}
    This condition is met for exceptional cutoff values, $\exlam$, for which the short-range part of the \LO{} wave function conspires to make the rows of the matrix in \cref{eq:A_3P0_in_text} linearly dependent~\cite{Gasparyan:2022isg}. As a consequence, the LECs $D^{(1)}_{^3P_0}$ and $E^{(0)}_{^3P_0}$ diverge, and the renormalization conditions can no longer be satisfied. In contrast to the former case, these divergences propagate to observables that acquire an unphysical cutoff dependence in the vicinity of $\exlam$.
\end{enumerate}

The two lower panels of \cref{fig:LECs_3P0} display the $^3P_0$ LECs at \NNLO{} as a function of the cutoff in the range relevant for the triton calculations in the following section. The solid vertical line shows the location of the limit-cycle-like cutoff at $\Lambda=679$~MeV, while the dashed vertical line shows the location of an exceptional cutoff, $\exlam = 670$~MeV. The short-distance part of the \LO{} wave function oscillates increasingly for greater cutoff values, leading to a repeating pattern of limit-cycle-like and exceptional cutoffs \cite{Gasparyan:2022isg}. Note that the exact locations at which the limit-cycle-like and exceptional cutoffs appear depend on the choices of regulator function and constants in the potentials.

Exceptional cutoffs and their relation to the \LO{} wave function is further studied in Appendix~\ref{app:1S0}, where a toy example (not including singular potentials) is presented. 

\begin{figure}
    \centering
    \includegraphics[width=\columnwidth]{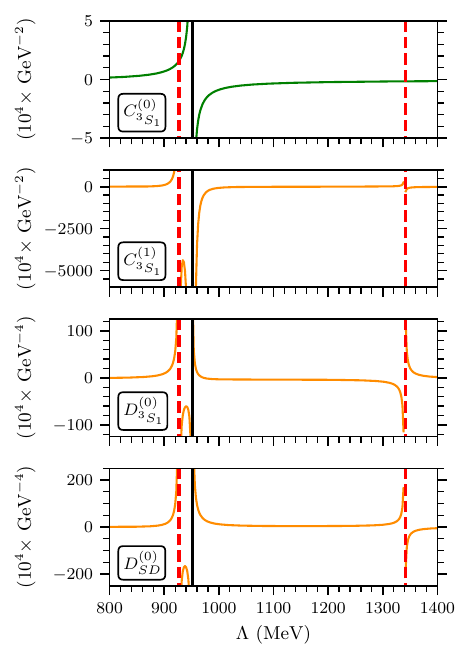}
    \caption{Values of the LECs at LO (top panel) and \NNLO{} (remaining panels) in the $\tS$ channel as a function of the cutoff, $\Lambda$. The vertical solid line indicates the location of a limit-cycle-like cutoff, while the vertical dashed lines indicate the location of exceptional cutoffs.}
    \label{fig:LECs_3S-D1}
\end{figure}

\subsection{The $\tS$ channel and the deuteron \label{sec:deutron}}

Let us continue to investigate exceptional cutoffs in the $\tS$ channel at \NNLO{} and their impact on the coupled-channel phase shifts and the deuteron ground-state energy. See Appendix~\ref{app:exceptional_cutoffs} for complete derivations as well as additional figures. Using notation analogous to the previously discussed $^3P_0$ case, the renormalization conditions used to determine the $\tS$ LECs at \NNLO{} are given by \cite{Thim:2024yks}
\begin{align}
    \delta^{(2)}_0(k_1) &= 0 \nonumber \\
    \delta^{(0)}_0(k_2) + \delta^{(2)}_0(k_2)  &= \delta_\mathrm{0,exp}(k_2) \label{eq:3S-D1_ren_cond}\\
    \epsilon^{(0)}(k_2) + \epsilon^{(2)}(k_2) &= \epsilon_\mathrm{exp}(k_2). \nonumber
\end{align}
Here, $k_1$ $(k_2)$ corresponds to $\Tl=30$~MeV $(\Tl=50$~MeV), and $\delta^{(\nu)}_0$ denotes a phase shift in the $\l=0$ channel while $\epsilon^{(\nu)}$ is the mixing angle at chiral order $\nu$. \Cref{eq:3S-D1_ren_cond} can be transformed into a matrix equation for the \NNLO{} LECs $\left(\vec{\alpha}^{(2)}\right)^T \equiv (C^{(1)}_{^3S_1},D^{(0)}_{^3S_1},D^{(0)}_{SD})$,
\begin{equation}
    A\vec{\alpha}^{(2)} = \vec{\delta},
    \label{eq:A_3S-D1_in_text}
\end{equation}
which is completely analogous to \cref{eq:A_3P0_in_text} (see also \cref{eq:A_3S-D1_app} for details).

\Cref{fig:LECs_3S-D1} shows the LECs in the $\tS$ channel as a function of the cutoff. It can be seen that the LO LEC exhibits a limit-cycle-like divergence, while the \NNLO{} LECs show additional divergences at the exceptional cutoffs. We have verified that the location of the limit-cycle-like and exceptional cutoffs agrees with the zeros of the determinant of $A$ in \cref{eq:A_3S-D1_in_text}. 

\begin{figure}
    \centering
    \includegraphics[width=\columnwidth]{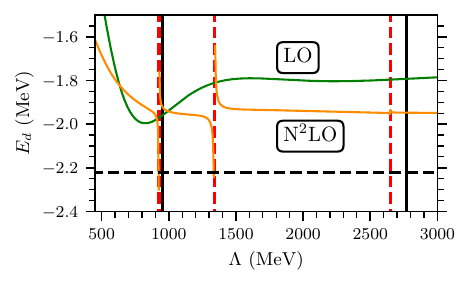}
    \caption{The ground-state energy of the deuteron, $E_d$, at \LO{} and \NNLO{} as a function of the cutoff $\Lambda$. The vertical solid lines indicate the locations of the limit-cycle-like cutoffs, while the vertical dashed lines indicate the location of the exceptional cutoffs. The horizontal dashed line shows the experimental value for $E_d$.}
    \label{fig:deuteron_Lambda}
\end{figure}

\begin{figure}
    \centering
    \includegraphics[width=\columnwidth]{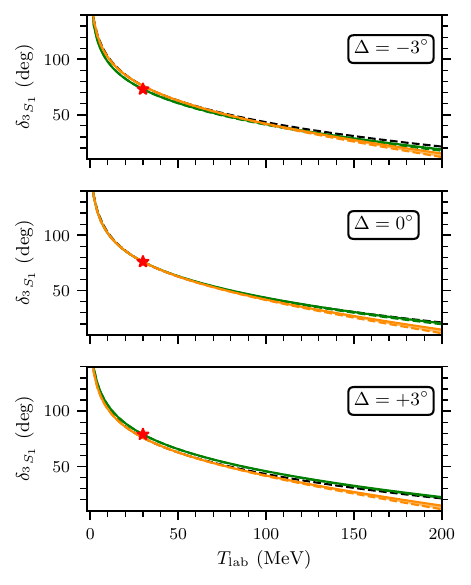}
    \caption{Phase shifts in the $^3S_1$ channel as a function of the laboratory scattering energy, $T_\mathrm{lab}$, for orders LO (green) and \NNLO{} (orange). The bands indicate the envelope of cutoff variation from $\Lambda=750$~MeV (dashed lines) to $\Lambda=1000$~MeV (solid lines). Note that the NLO contribution in $\tS$ is zero.  The red star shows the phase shift used to calibrate the LO LEC, which is shifted $-3^\circ$, $0^\circ$ and $+3^\circ$ compared to the Nijmegen \cite{Stoks:1993tb} phase shift in the top, middle, and bottom panel, respectively. The black dashed line shows the phase shift from the Nijmegen partial wave analysis \cite{Stoks:1993tb}.}
    \label{fig:3S1_shift_phases}
\end{figure}

Having analyzed the location of the exceptional cutoffs in $\tS$ we now study predictions for the ground-state energy of the deuteron, $E_d$, as a function of the cutoff. In \cref{fig:deuteron_Lambda} we show $E_d$ at \LO{} and \NNLO, since the \NLO{} contribution is zero in the $\tS$ channel (see \cref{tab:potentials_PC}). The \NNLO{} prediction diverges at the exceptional cutoffs, which is most clearly visible in the figure at $\Lambda = 1340$ MeV. However, note that there is no trace of the limit-cycle-like divergence.

It was demonstrated in Ref.~\cite{Gasparyan:2022isg} that exceptional cutoffs in the $^3P_0$ channel typically occur at values slightly below a corresponding limit-cycle–like cutoff. In the coupled $\tS$ channel, we observe a slightly different pattern, and the exceptional cutoff does not always appear directly below an associated limit-cycle-like cutoff. One such example is the exceptional cutoff at $\Lambda=1340$~MeV. We also note that the pattern of exceptional cutoffs in the $\tS$ channel is consistent with the LECs reported in~\cite{Shi:2022blm}.

We will now study the exceptional cutoff at $\Lambda = 930$~MeV in more detail. This is also the one most relevant for the upcoming triton calculations. Following Ref.~\cite{Peng:2024aiz}, we exploit the freedom within the EFT to slightly modify the \LO{} potential to temper the divergence in $E_d$ at this exceptional cutoff. Such modifications are consistent with EFT principles, provided they can be perturbatively corrected at subleading orders. In practice, we modify the potential by shifting the calibration datum for the $^3S_1$ phase shift at $T_\mathrm{lab} = 30$~MeV. Thus, we define a new \LO{} calibration phase shift as
\begin{equation}
    \delta_{^3S_1,\mathrm{exp}}(k_1) \to  \delta_{^3S_1,\mathrm{exp}}(k_1) + \Delta. \ 
    \label{eq:3S1_phase_shift_shift}
\end{equation}
In \cref{fig:3S1_shift_phases} we show the $^3S_1$ phase shifts resulting from different choices of $\Delta$. While the \LO{} phase shift is affected by $\Delta$, this dependence is perturbatively corrected at \NNLO{} (see also \cref{fig:3S-D1_shift_phases} in Appendix~\ref{app:exceptional_cutoffs} for the remaining coupled-channel phase shifts).
\begin{figure}
    \centering
    \includegraphics[width=\columnwidth]{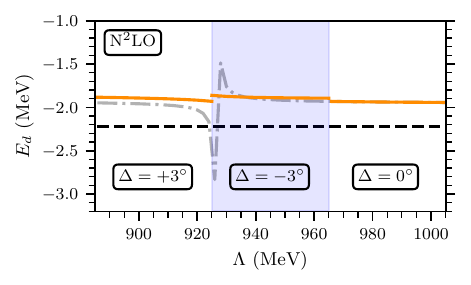}
    \caption{Predicted deuteron ground-state energy at \NNLO{} as a function of $\Lambda$ for different values of the shift $\Delta$. The solid lines in each cutoff region correspond to shifts $\Delta=+3^\circ,-3^\circ$ and $0^\circ$, respectively, as defined in \cref{eq:3S1_phase_shift_shift}. The dashed-dotted line shows the result for $\Delta=0^\circ$, and the dashed line shows the experimental value for $E_d$.
    }
    \label{fig:deuteron_shifts}
\end{figure}

In \cref{fig:deuteron_shifts} we show the predicted deuteron ground-state energy at \NNLO{} with suitably chosen shifts, $\Delta=-3^\circ,+3^\circ$ and $0^\circ$ in different cutoff regions. By altering the calibration phase shift, we modify the \LO{} potential and the resulting \LO{} wave function. This results in the exceptional cutoff being shifted towards lower (higher) values for a negative (positive) shift $\Delta$. By imposing a $\Lambda$-dependent shift $\Delta$, the exceptional divergence in $E_d$ can be avoided altogether at \NNLO{}. The shift is applied only when computing predictions beyond \LO. Moreover, since an \NNLO{} prediction is obtained as the sum of the \LO{}, \NLO{}, and \NNLO{} contributions, the effects of $\Delta$ at \LO{} and \NLO{} are not meaningful separately. Note that the unshifted LO prediction is the one shown in \cref{fig:deuteron_Lambda}.

\section{The three-nucleon system up to \NNLO \label{sec:triton}}
\begin{figure}
    \centering
    \includegraphics[width=0.7\columnwidth]{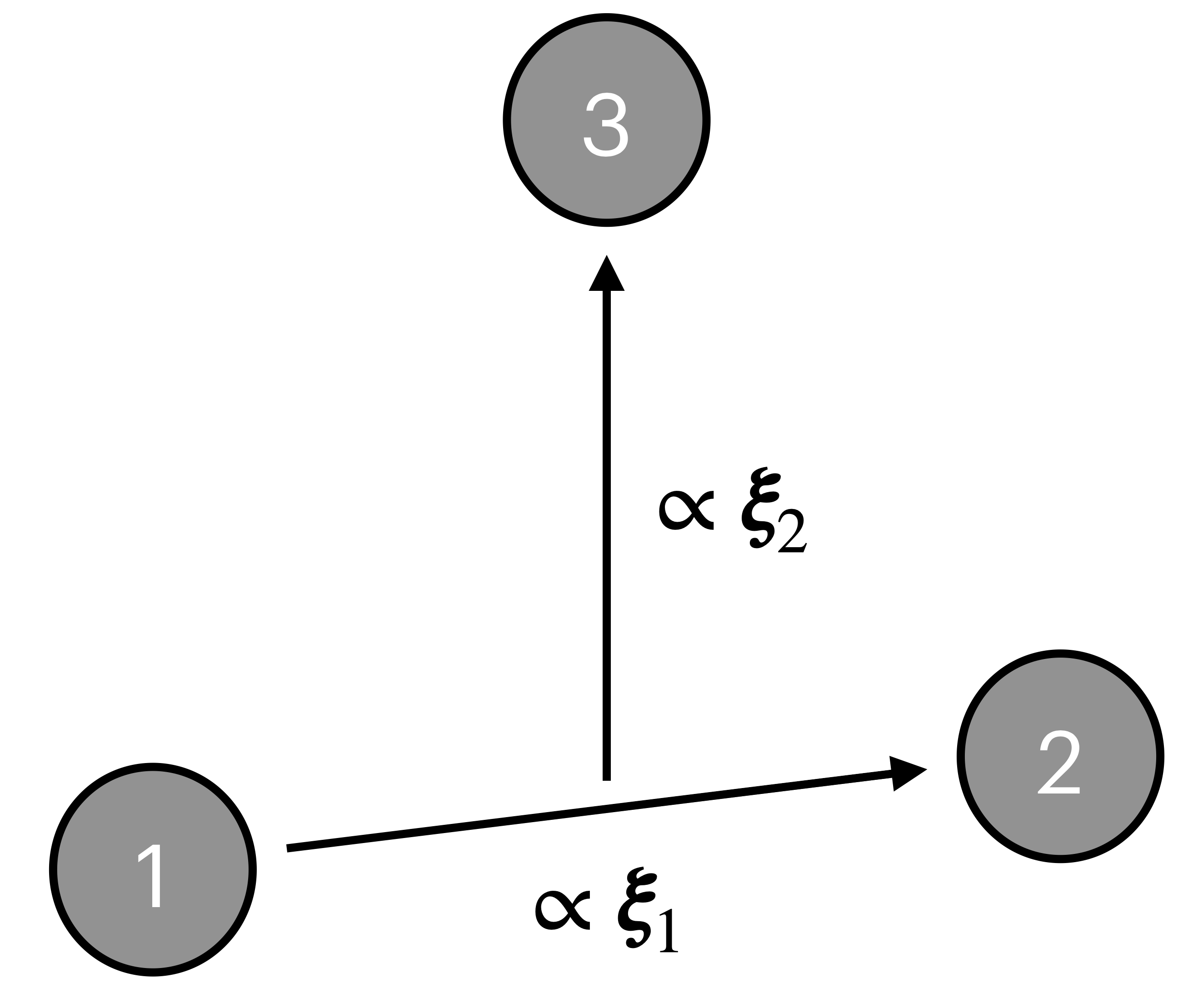}
    \caption{A system of three nucleons described by Jacobi coordinates $\bm{\xi}_1$ and $\bm{\xi}_2$.}
    \label{fig:3N-system}
\end{figure}

In this section, we predict and analyze the ground-state energy of the triton up to \NNLO{} within the Long and Yang PC scheme. To solve the time-independent Schrödinger equation for the three-nucleon system, we employ the NCSM, formulated in relative (Jacobi) coordinates~\cite{Navratil:1999pw}, as illustrated in \cref{fig:3N-system}. The NCSM Python code that we developed for this purpose is publicly available~\cite{py-ncsm}. 

\subsection{The NCSM in Jacobi coordinates for the triton}

The three-nucleon system can be described using basis states
\begin{equation}
    \ket{n\l m_\l} \otimes \ket{\N \L m_\L} \otimes  \ket{(s \frac{1}{2}) S m_S} \otimes \ket{(t\frac{1}{2})T m_T},
\end{equation}
where eigenfunctions of the three-dimensional spherical harmonic oscillator (HO) $\ket{n\l m_{\l}}$ and $\ket{\N\L m_{\L}}$ form a basis for the Jacobi-coordinate states $\ket{\bm{\xi}_1}$ and $\ket{\bm{\xi}_2}$, respectively. The oscillator length, $b$, entering these eigenfunctions is related to the HO frequency, $\omega$, as $b=(\mN \omega)^{-1/2}$. 
Moreover, $s$ and $t$ denote the coupled spin and isospin of nucleons 1 and 2 while $S$ and $T$, with associated projections $m_S$ and $m_T$, denote the total spin and isospin for the two-nucleon subsystem coupled with the third nucleon. 

With coupled angular momenta, more convenient three-nucleon basis states are formed as
\begin{align}
    &\ket{\alpha;JT}_{12} \equiv \ket{n \l s j t, \N \L \J; JT} \nonumber \\ &\equiv \ket{n\N (\l s) j (\L \frac{1}{2}) \J; (j\J)J m_J}  \otimes \ket{(t\frac{1}{2}) T m_T}.
    \label{eq:Nalpha}
\end{align}
The quantum numbers $m_J$ and $m_T$ are dropped by rotational symmetry and assumed isospin symmetry, respectively, and we use $\alpha\equiv(n \N \l s j t \L \J)$.
Here, $j$ denotes the total angular momentum of the relative system of nucleons 1 and 2, while $\J$ is the total angular momentum of nucleon 3 relative to the \CM{} of nucleons 1 and 2. Furthermore, $J$ denotes the total angular momentum of the three-nucleon system. The two-nucleon subsystem is antisymmetric by the condition $(-1)^{\l + s + t} = -1$. Hence, the basis states in \cref{eq:Nalpha} are only partially antisymmetric. 

Fully antisymmetric basis states are constructed by diagonalizing the three-particle antisymmetrizer in the partially antisymmetric basis, following Ref.~\cite{Navratil:1999pw}. This antisymmetrizer is diagonal in the quantum numbers $(N,J,T)$, where $N \equiv 2n + \l + 2\N + \L$. Fully antisymmetric states can be expressed as
\begin{align}
    &\ket{NJT, \gamma}  \nonumber \\&= \sum_{\alpha \ : \ \substack{2n+\l + \\2\N + \L = N}} {}_{12}\braket{\alpha; JT | NJT,\gamma} \ket{\alpha; JT}_{12}
    \label{eq:Gamma}
\end{align}
where $\gamma$ enumerates states of fixed $(N,J,T)$. The complete set of fully antisymmetric states, $\{\ket{\Gamma}\}$, for a specific spin, parity, and isospin $(J^\Pi, T)$ can now be obtained from \cref{eq:Gamma}. The model-space basis is truncated by total HO energy, i.e., only allowing states with $N\leq N_\mathrm{max}$.

\subsection{The triton at leading order}
The \LO{} three-nucleon Hamiltonian in the \cm{} frame reads
\begin{equation}
      H^{(0)}_{\mathrm{\cm}} = \frac{1}{3}\sum_{i<j=1}^3 \frac{(\bm{p}_i-\bm{p}_j)^2}{2\mN} + \sum_{i<j=1}^3 V^{(0)}_{ij}.
      \label{eq:Hcm}
\end{equation}
Here, $\mN$ denotes the nucleon mass, $\bm{p}_i$ the $i$:th nucleon momentum in the \CM{}~frame, and $V^{(0)}_{ij}$ the \LO{} two-nucleon potential for nucleons $i$ and $j$. We neglect the mass difference between the proton and the neutron \cite{Kamuntavicius:1999eu}. 

We are interested in the triton ground state and construct basis states with $(J^\Pi, T) = (\frac{1}{2}^+, \frac{1}{2})$ according to \cref{eq:Gamma}. The Schrödinger equation is solved by diagonalizing the \LO{} Hamiltonian in this fully antisymmetric basis
\begin{equation}
    \sum_{\Gamma'}\bra{\Gamma} H^{(0)}_{\mathrm{\cm}}\ket{\Gamma'}\braket{\Gamma'|\Psi^{(0)}_n} = E^{(0)}_n \braket{\Gamma|\Psi^{(0)}_n},
\end{equation}
and the full spectrum $\{\ket{\Psi^{(0)}_k},E^{(0)}_k\}_k$ can be obtained. We employ HO frequencies $\omega \in [10 \ \mathrm{MeV},125 \ \mathrm{MeV}]$---where the large values are needed for computations with high-momentum cutoffs---and consider basis truncations up to $\Nmax=46$. The size of the partially- and fully-antisymmetric basis for $\Nmax=46$ is $19000$ and $6336$, respectively. This means that exact diagonalization of the full Hamiltonian is possible and that the complete spectrum of states can be obtained. Knowledge of the full \LO{} spectrum will be important when we compute the \NNLO{} corrections in perturbation theory.

\begin{figure}
    \centering
    \includegraphics[width=\columnwidth]{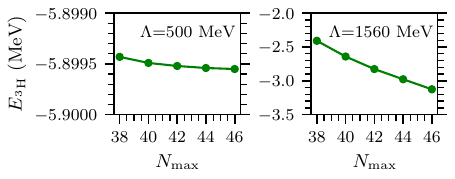}
    \caption{The triton ground-state energy at \LO{} for $\Lambda=500$~MeV (left panel), and $\Lambda=1560$~MeV (right panel). The HO frequency of the variational minimum is chosen for each $\Nmax$.}
    \label{fig:Nmax-convergence}
\end{figure}

The convergence of the triton ground-state energy is shown in \cref{fig:Nmax-convergence} as a function of $\Nmax$ for different cutoffs. In this work we consider cutoffs up to $\Lambda=1560$~MeV. As the cutoff increases, it becomes increasingly challenging to obtain converged results.

\begin{figure}
    \centering
    \includegraphics[width=0.99\columnwidth]{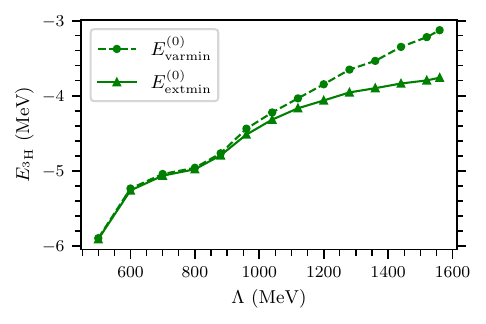}
    \caption{Triton ground-state energy at \LO{} as a function of the momentum cutoff, $\Lambda$. The dashed (solid) lines show the variational (infrared extrapolated) energies.}
    \label{fig:LO-ext}
\end{figure}

Our strategy to extend the reach of NCSM calculations to large cutoff values combines the use of bases with large HO frequencies (to handle ultraviolet physics) with an infrared extrapolation technique \cite{Forssen:2017wei}. This implies that computations are mainly performed for HO frequencies higher than those corresponding to the location of the variational minimum, and that the NCSM basis parameters are translated to relevant infrared and ultraviolet scales~\cite{Wendt:2015nba}. Infrared-extrapolated \LO{} ground-state energies, $E^{(0)}_\mathrm{extmin}$, are shown in \cref{fig:LO-ext} together with the variational minimum for $\Nmax=46$, $E^{(0)}_\mathrm{varmin} \equiv \mathrm{min}_\omega \ E^{(0)}(\omega)$. The extrapolated energy appears to reach a plateau when $\Lambda$ is increased, consistent with the findings of a similar study in Ref.~\cite{Song:2016ale} using the Faddeev equations. 

The obtained ground-state energy shown in \cref{fig:LO-ext} was found to be sensitive to the calibration procedure used to fix the LECs. We observed a variation of the triton ground-state energy of $\approx 1$~MeV if we, instead of low-energy phase shifts, calibrate the two $S$-wave LECs to the singlet scattering length and the deuteron ground-state energy. This is consistent with the findings reported in Ref.~\cite{Song:2016ale} and suggests that a more robust calibration procedure is required to draw quantitative conclusions about the quality of the predictions---a task we leave for future work.

\begin{figure*}
    \centering
    \includegraphics[width=0.84\textwidth]{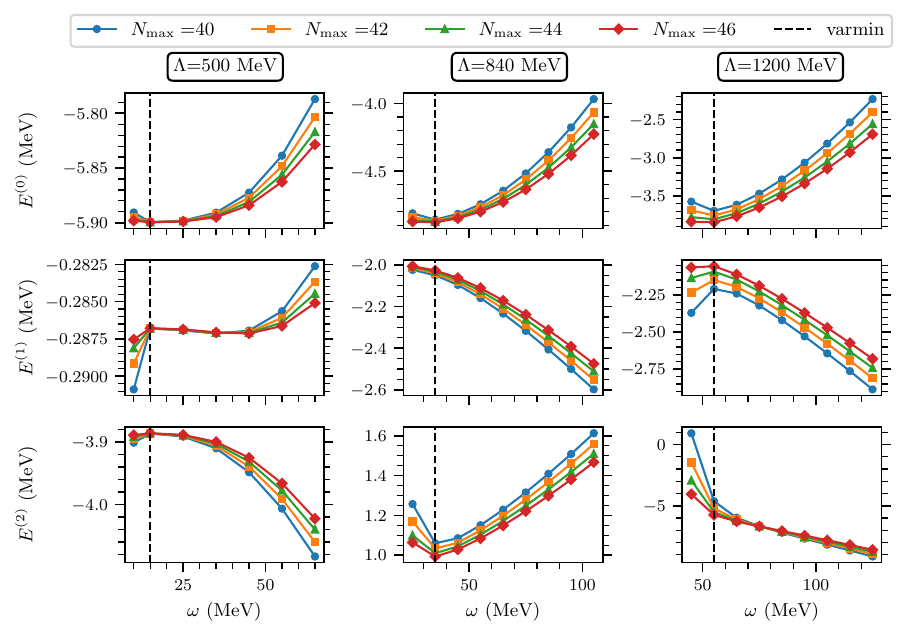}
    \caption{Contributions to the triton ground-state energy at, LO (top row), NLO (middle row) and \NNLO{} (bottom row) as a function of the HO frequency, $\omega$, for different model space truncations $\Nmax$. The columns show results for three different cutoffs, $\Lambda$. The vertical dashed lines indicate the LO variational minimum. Note that the $\omega$-interval varies with the cutoff.}
    \label{fig:E_hw_Nmax}
\end{figure*}

Spurious deeply bound \NN{} states appear in the $^3P_0$ and $\tS$ channels as the LO LECs exhibit a limit-cycle-like divergence, i.e., for cutoffs $\Lambda > 679$~MeV. These spurious states need to be projected out from the \NN{} spectrum before performing triton calculations. We do this using the technique of orthogonalizing pseudopotentials ~\cite{Kukulin:1978he,Song:2016ale,Nogga:2005hy}. This is implemented by transforming the LO potential as
\begin{equation}
    V^{(0)} \to  V^{(0)} + \sum_\phi \lambda_\phi \ket{\phi}\bra{\phi},
    \label{eq:OPP}
\end{equation}
where $\ket{\phi}$ denotes the spurious \NN{} states and $\lambda_\phi$ a large positive constant of order $10^8$~MeV . This procedure removes the effect of the spurious states in the triton wave functions at \LO. It turns out that this projection is sufficient to also remove the effect of the spurious states in subleading corrections, which is verified in the following subsection.

\subsection{The triton at subleading orders}

Subleading corrections to the triton ground-state energy are computed in perturbation theory, where the \NLO{} contribution reads 
\begin{equation}
    E^{(1)} = \braket{\Psi^{(0)}_0 |\sum_{i<j=1}^3 V^{(1)}_{ij}|\Psi^{(0)}_0}.
    \label{eq:E_NLO}
\end{equation}
In this work we are only interested in the ground-state energy. Therefore, only the \LO{} ground-state wave function, $\ket{\Psi^{(0)}_{0}}$, is needed to compute the \NLO{} perturbation.

The \NNLO{} contribution to the ground-state energy is 
\begin{align}    
 E^{(2)} &= \braket{\Psi^{(0)}_0|\sum_{i<j=1}^3 V^{(2)}_{ij}|\Psi^{(0)}_0} \nonumber \\ &+ \sum_{m\neq 0} \frac{|\braket{\Psi^{(0)}_0 |\sum_{i<j=1}^3 V^{(1)}_{ij}|\Psi^{(0)}_m}|^2}{E^{(0)}_0-E^{(0)}_m},
 \label{eq:E_NNLO}
\end{align}
where two insertions of the \NLO{} potential and one insertion of the \NNLO{} potential contribute. Note also that the \NNLO{} correction to the ground state requires the full LO spectrum $\{\ket{\Psi^{(0)}_m},E^{(0)}_m\}_m$. Although the energy denominator $(E^{(0)}_0-E^{(0)}_m)$ suppresses highly excited states in \cref{eq:E_NNLO}, we find that all states must be considered to obtain a converged result.

Perturbative corrections from NN channels with LO spurious states  ($^3P_0$ and $\tS$) first enter in the expectation value of the \NNLO{} potential in \cref{eq:E_NNLO}. We confirm that this expectation value is free from the interference of spurious states by numerically verifying the relation
\begin{align}
    \braket{\Psi^{(0)}_k|\sum_{i<j=1}^3 (1-P_\phi)_{ij} V^{(2)}_{ij}(1-P_\phi)_{ij}|\Psi^{(0)}_k} \nonumber \\ \to_{\lambda_\phi \to \infty} \braket{\Psi^{(0)}_k|\sum_{i<j=1}^3 V^{(2)}_{ij}|\Psi^{(0)}_k},
    \label{eq:spurious_N2LO}
\end{align}
for sufficiently high values of $\lambda_\phi$. The projectors for the $ij$ subsystems are defined as $(1-P_\phi)_{ij} = (1-\sum_\phi \ket{\phi}\bra{\phi})_{ij}$. 

We compute \NLO{} and \NNLO{} contributions to the triton ground-state energy from \cref{eq:E_NLO,eq:E_NNLO} for different values of $\omega$ and $\Nmax$ and for cutoffs values in the range $500$~MeV~$\leq \Lambda\leq 1560$~MeV. Results for a few representative cutoff values are shown in \cref{fig:E_hw_Nmax}. We immediately notice that the variational principle does not hold beyond LO, as expected for a perturbative calculation. In particular, $E^{(2)}$ in the lower-left panel shows a concave parabolic behavior as a function of $\omega$ for all studied values of $\Nmax$. This complicates the selection of $\omega$ for assessing the convergence. 
We observe that the convergence in $\Nmax$ becomes slower as the chiral order and cutoff increase. Nevertheless, within the range of $\Nmax$ considered, the perturbative corrections show a decreasing dependence on $\omega$, indicating the expected convergence with increasing basis size. As a result, we obtain reasonably converged results at \NNLO{} for $\Lambda \lesssim 1200$~MeV.

\begin{figure}
    \centering
    \includegraphics[width=0.99\columnwidth]{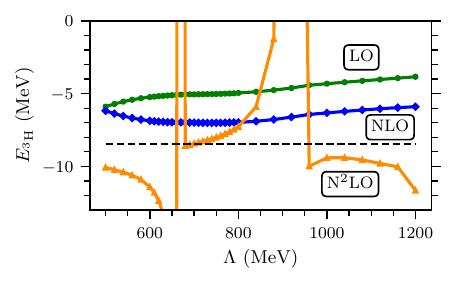}
    \caption{The ground-state energy of the triton up to \NNLO{} as a function of the momentum cutoff, $\Lambda$, computed in the NCSM. The HO frequency of the LO variational minimum for $\Nmax=46$ is used at all orders, see \cref{fig:E_hw_Nmax}. The black dashed line shows the experimental value \cite{Purcell:2010hka}.}
    \label{fig:fN2LO_triton_varmin}
\end{figure}

\Cref{fig:fN2LO_triton_varmin} shows the order-by-order predictions for the ground-state energy of the triton, i.e.,
\begin{equation}
    E_{\mathrm{N}^\nu \mathrm{LO}} = \sum_{\beta = 0}^\nu E^{(\beta)}.
\end{equation}
At each cutoff, we use the HO frequency of the LO variational minimum for all orders and do not include any IR extrapolations at LO. Our results at LO and NLO are similar to previous studies \cite{Song:2016ale,Yang:2020pgi}. For the new \NNLO{} result, however, we observe a strong cutoff dependence in the vicinities of $\Lambda \approx650$~MeV and $\Lambda\approx900$~MeV. This cutoff dependence will now be investigated in more detail.

\begin{figure}
    \centering
    \includegraphics[width=0.99\columnwidth]{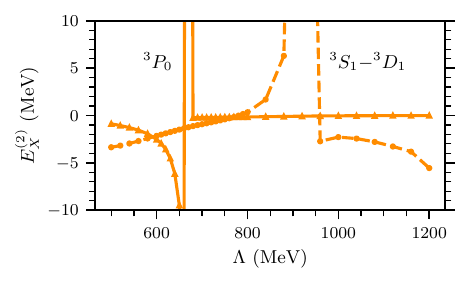}
    \caption{Channel contributions, $E^{(2)}_X$, for $X = {}^3P_0$ (solid line) and $X={}\tS$ (dashed line) computed using \cref{eq:E_NNLO_proj}.}
    \label{fig:N2LO_triton_chn_cont}
\end{figure}

\subsection{The effect of exceptional cutoffs in the triton}

To understand the divergent behavior of the triton ground-state energy at \NNLO, it is instructive to study contributions from specific \NN{} channels. We define the contribution at NLO and \NNLO{} from different \NN-channels, $X$, as
\begin{align}
    E^{(1)}_X &\equiv \braket{\Psi^{(0)}_0 |  \sum_{i<j=1}^3 P_X V^{(1)}_{ij} P_X|\Psi^{(0)}_0}, 
    \label{eq:E_NLO_proj} \\
    E^{(2)}_X &\equiv \braket{\Psi^{(0)}_0|\sum_{i<j=1}^3 P_X V^{(2)}_{ij} P_X|\Psi^{(0)}_0} \nonumber \\ &+ \sum_{m\neq 0} \frac{|\braket{\Psi^{(0)}_0 |\sum_{i<j=1}^3 P_X V^{(1)}_{ij} P_X|\Psi^{(0)}_m}|^2}{E^{(0)}_0-E^{(0)}_m}.
    \label{eq:E_NNLO_proj}
\end{align}
Here, $P_X$ denotes a projector onto an \NN{} channel $X\in\mathcal{C}_\mathrm{NN}=\{^1S_0,\ ^3P_0,\ ^1P_1,\ ^3P_1,\ \tS,\dots\}$.

The NLO expectation value is linear in the projector and the decomposition becomes additive
\begin{equation}
       E^{(1)} = \sum_{X \in \mathcal{C}_\mathrm{NN}} E^{(1)}_X
\end{equation}
As such, the contribution from a given \NN{} channel is well-defined. However, the second term in \cref{eq:E_NNLO_proj} is not linear in the projections, so defining a contribution from a given channel depends on how cross terms are treated, and as such the decomposition is ambiguous. However, if one removes $V^{(0)}_{1\pi}$ from the NLO potential (see \cref{tab:potentials_PC}) only the $^1S_0$ channel would have a non-zero contribution in the second term, and unambiguous channel projections are possible. Our computations show that  $V^{(0)}_{1\pi}$ only yields a percent-level contribution to the second term of \cref{eq:E_NNLO_proj} such that approximate additivity also holds at \NNLO{}
\begin{equation}
       E^{(2)} \approx \sum_{X \in \mathcal{C}_\mathrm{NN}} E^{(2)}_X.
\end{equation}

Channels of particular interest for this study are $^3P_0$ and $\tS$ for which we have observed exceptional cutoff values. In \cref{fig:N2LO_triton_chn_cont} we show the \NNLO{} contributions, $E^{(2)}_{X}$, for these channels. These results confirm that the first divergence, at $\Lambda \approx 650$~MeV, originates from the $^3P_0$ channel, while the second, at $\Lambda \approx 900$~MeV, originates from the $\tS$ channel. In the latter case, we also observe a broad domain of cutoffs with $\Lambda < \exlam$ over which the triton ground-state energy is impacted. We will now investigate whether these divergences can be avoided by shifting the LO wave function using the same approach as in the \NN{} sector explored in Sec.~\ref{sec:exc-nn}.

\subsection{Shifting the LO triton wave function}

\begin{figure}
    \centering
    \includegraphics[width=\columnwidth]{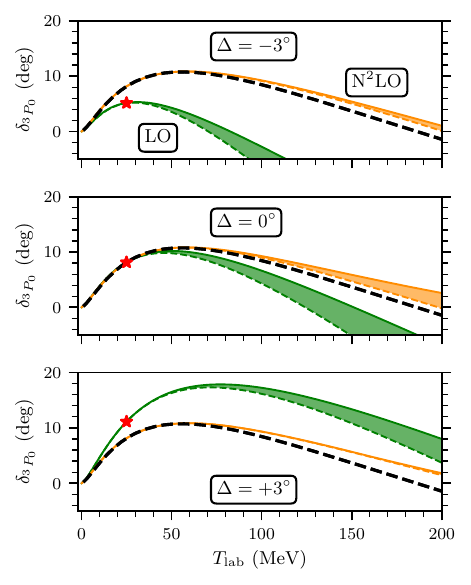}
    \caption{Phase shifts in the $^3P_0$ channel as a function of the laboratory scattering energy, $T_\mathrm{lab}$, for orders LO (green) and \NNLO{} (orange). Note that the NLO contribution in $^3P_0$ is zero. The bands indicate the envelope of cutoff variation from $\Lambda=500$~MeV (dashed lines) to $\Lambda=790$~MeV (solid lines). The red star indicates the LO phase shift used to calibrate the LO LEC that is shifted $-3^\circ$, $0^\circ$ and $+3^\circ$ compared to the Nijmegen \cite{Stoks:1993tb} phase shift in the top, middle, and lower panel, respectively. The black dashed line shows the phase shift from the Nijmegen partial wave analysis \cite{Stoks:1993tb}. }
    \label{fig:3P0_shift_phases}
\end{figure}

Nominally, we calibrate the LO $^3P_0$ LEC to the Nijmegen phase shift at $T_\mathrm{lab} = 25$~MeV \cite{Thim:2024yks}. In an attempt to resolve the exceptional cutoff divergence at $\Lambda \approx 650$~MeV, we make use of the EFT freedom to modify the LO potential, following the approach outlined in Sec.~\ref{sec:deutron} \cite{Peng:2024aiz}. We define the LO calibration data point by slightly moving the value for the Nijmegen phase shift 
\begin{equation}
    \delta_{^3P_0,\mathrm{exp}} \to  \delta_{^3P_0,\mathrm{exp}} + \Delta,
    \label{eq:3P0_phase_shift_shift}
\end{equation}
where we introduce a shift $\Delta \in [-3^\circ,3^\circ]$. \Cref{fig:3P0_shift_phases} shows the $^3P_0$ phase shifts for three different values of $\Delta$. Clearly, the LO phase shift varies considerably for the different $\Delta$'s while the \NNLO{} phase shift does not. This observation indicates that the effect of the arbitrary shift can be perturbatively corrected at higher orders.

\begin{figure}
    \centering
    \includegraphics[width=\columnwidth]{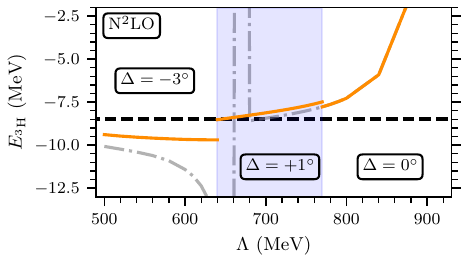}
    \caption{The triton ground-state energy at \NNLO{} as a function of the momentum cutoff. The colored solid lines in each cutoff region correspond to shifts $\Delta=-3^\circ,+1^\circ$ and $0^\circ$, respectively, as defined in \cref{eq:3P0_phase_shift_shift} for the $^3P_0$ channel. The gray dashed-dotted line shows the $\Delta=0^\circ$ result across all cutoffs. The black dashed line shows the experimental triton ground-state energy \cite{PDG_2022}. All calculations use $N_\mathrm{max}=46$ and the HO frequency, $\omega$, is chosen as the LO variational minimum at all orders.}
    \label{fig:triton_shifts}
\end{figure}

\Cref{fig:triton_shifts} shows predicted triton energies up to \NNLO{} using \LO{} potentials with shifts $\Delta=-3^\circ,+1^\circ$ and $0^\circ$. The gray dashed-dotted line shows the computation for $\Delta=0^\circ$ across all cutoffs, where the exceptional divergence is visible. The location of the exceptional cutoff is shifted towards lower (higher) cutoffs for a negative (positive) shift $\Delta$. Consequently, by choosing different values for $\Delta$ in distinct cutoff regions, a finite prediction for the triton ground-state energy can be achieved at \NNLO{}. As discussed in connection with \cref{fig:deuteron_shifts}, the shift in the LO potential is only introduced when computing the \NNLO{} prediction. The \LO{} and \NLO{} predictions are still obtained with $\Delta = 0^\circ$ for all cutoffs.

\begin{figure}
    \centering
    \includegraphics[width=\columnwidth]{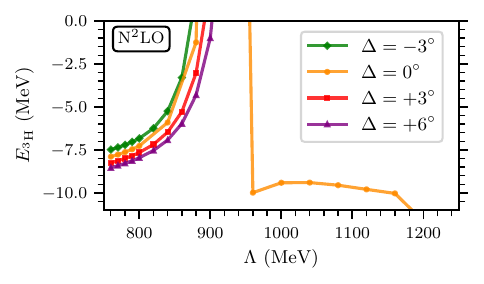}
    \caption{Triton ground-state energy at \NNLO{} as a function of the cutoff. The lines with different markers show the predicted energy for different shifts, $\Delta$, in $\tS$ channel (see \cref{eq:3S1_phase_shift_shift}).}
    \label{fig:triton_3S1_shift}
\end{figure}

For $\Lambda \gtrsim 800$~MeV, the effect of the exceptional cutoff in the $\tS$ channel needs to be addressed. In Sec.~\ref{sec:deutron} we demonstrated how to modify the \LO{} wave function in this channel to resolve the divergence caused by the exceptional cutoff for the deuteron ground-state energy. When applied to the triton, the results are less encouraging. In \cref{fig:triton_3S1_shift} we show the triton ground-state energy at \NNLO{} for different shifts, $\Delta \in [-3^\circ,6^\circ]$, as defined in \cref{eq:3S1_phase_shift_shift}. The effect of applying the shift is smaller compared to the $^3P_0$ channel. Although the location of the exceptional cutoff moves slightly, the change is not sufficient when compared to the range of cutoff values over which the divergence impacts the triton ground-state energy. Therefore, the strategy of applying different shifts in separate cutoff domains to remedy the effects of the exceptional cutoff does not appear to be effective in this case. 

There is the possibility of exploring larger shifts in the $\tS$ channel since the relative size of the LO shift is much smaller in $\tS$ compared to $^3P_0$, see \cref{fig:3S1_shift_phases,fig:3P0_shift_phases}. However, for shifts $\Delta \gtrsim 6^\circ$  or $\Delta \lesssim -3^\circ$, the obtained LO deuteron ground-state energy cannot be perturbatively corrected at \NNLO{}. Thus, we do not see this as a viable approach to avoid the divergence observed in \cref{fig:triton_3S1_shift}.

\section{Conclusions and Outlook \label{sec:summary}}
In this work, we have taken the first steps of using the PC by Long and Yang in the few-nucleon sector up to \NNLO{} in perturbation theory. We employed the NCSM to perform perturbative computations for the triton ground-state energy, and obtained converged results for cutoffs $\Lambda \lesssim 1200$~MeV. We have reviewed how so-called exceptional cutoffs arise in the \NN{} system in this PC \cite{Gasparyan:2022isg,Peng:2024aiz} and extended previous work by considering also the coupled $\tS$ channel. 
We predicted the triton ground-state energy up to \NNLO{} and interpreted the results using our understanding of exceptional cutoff values. 

The conclusions can be summarized as follows.

\begin{enumerate}
    \renewcommand{\labelenumi}{(\roman{enumi})}
    \item We show that exceptional cutoffs also arise in the $\tS$ channel, where the pattern of appearance is slightly different compared to the $^3P_0$ channel. In particular, exceptional cutoffs appear at more irregular locations and not just below an associated limit-cycle-like cutoff, see \cref{fig:LECs_3S-D1}. For example, one exceptional cutoff appears at $\Lambda\approx 1340$~MeV, whose effect is observed in the triton calculations, see \cref{fig:fN2LO_triton_varmin}.
    
    \item The predicted deuteron ground-state energy at \NNLO{} diverges at the exceptional cutoffs, see \cref{fig:deuteron_Lambda}. We apply the method proposed in Ref.~\cite{Peng:2024aiz} to shift the LO wave function and successfully remove the effect of the divergences, see \cref{fig:deuteron_shifts}.
    
    \item We can obtain converged perturbative computations of the triton ground-state energy using the NCSM up to the cutoff $\Lambda\approx1200$~MeV. However, effects of exceptional cutoffs in the $^3P_0$ and $\tS$ channels propagate to the triton \NNLO{} result producing divergences, see \cref{fig:fN2LO_triton_varmin} and \cref{fig:N2LO_triton_chn_cont}.
    
    \item We demonstrate that the method proposed in Ref.~\cite{Peng:2024aiz} effectively removes the exceptional cutoff divergence originating from the $^3P_0$ channel in the triton at \NNLO, as shown in \cref{fig:triton_shifts}.
    However, we find that the same method is not well-suited for taming the effect of the exceptional cutoff stemming from the $\tS$ channel, see \cref{fig:triton_3S1_shift}.
\end{enumerate}

The appearance of exceptional cutoffs seems to be intimately related to applying perturbation theory using a LO wave function with uncontrolled short-distance behavior. A recent study systematically investigated the connection between exceptional cutoffs and regulators \cite{PavonValderrama:2025zzk}. The findings suggest that exceptional cutoffs can be avoided by using a suitable type of regulation---which remains to be studied further.

Furthermore, the order-by-order convergence of few-nucleon observables should be investigated in more detail. This requires an improved method of inferring the LECs such that the EFT error can be properly assessed to facilitate more robust predictions.
The effect of including three-nucleon forces \cite{Weinberg:1992yk,vanKolck:1994yi} should also be investigated.

Exceptional cutoffs currently present a challenge for reliably computing nuclear observables beyond \NLO{} for $\Lambda \gtrsim 650$~MeV. However, as seen in our deuteron results, there are wide regions at $\Lambda \gtrsim 1500$~MeV without exceptional cutoffs, implying that the divergences can be avoided in principle. This region remains challenging to access in many-body computations. There are fortunately no exceptional cutoffs in the vicinity of $\Lambda \approx 500$~MeV where many-body computations can be converged. Studying heavier-mass nuclei at relatively low cutoffs therefore provides an opportunity to analyze the predictive differences between perturbative schemes (guided by RG invariance) and the non-perturbative WPC commonly employed in \chEFT.

\begin{acknowledgments}
O.T.~acknowleges interesting and stimulating discussions during the workshop "The nuclear interaction: post-modern developments" at ECT$^*$ Aug 19–23, 2024. This work was supported by the Swedish Research Council (Grants No.~2020-05127, No.~2021-04507, and No.~2024-04681). 
\end{acknowledgments}
\FloatBarrier

\bibliography{bib} 

\providecommand{\noopsort}[1]{}\providecommand{\singleletter}[1]{#1}%
\begin{thebibliography}{56}%
\makeatletter
\providecommand \@ifxundefined [1]{%
 \@ifx{#1\undefined}
}%
\providecommand \@ifnum [1]{%
 \ifnum #1\expandafter \@firstoftwo
 \else \expandafter \@secondoftwo
 \fi
}%
\providecommand \@ifx [1]{%
 \ifx #1\expandafter \@firstoftwo
 \else \expandafter \@secondoftwo
 \fi
}%
\providecommand \natexlab [1]{#1}%
\providecommand \enquote  [1]{``#1''}%
\providecommand \bibnamefont  [1]{#1}%
\providecommand \bibfnamefont [1]{#1}%
\providecommand \citenamefont [1]{#1}%
\providecommand \href@noop [0]{\@secondoftwo}%
\providecommand \href [0]{\begingroup \@sanitize@url \@href}%
\providecommand \@href[1]{\@@startlink{#1}\@@href}%
\providecommand \@@href[1]{\endgroup#1\@@endlink}%
\providecommand \@sanitize@url [0]{\catcode `\\12\catcode `\$12\catcode
  `\&12\catcode `\#12\catcode `\^12\catcode `\_12\catcode `\%12\relax}%
\providecommand \@@startlink[1]{}%
\providecommand \@@endlink[0]{}%
\providecommand \url  [0]{\begingroup\@sanitize@url \@url }%
\providecommand \@url [1]{\endgroup\@href {#1}{\urlprefix }}%
\providecommand \urlprefix  [0]{URL }%
\providecommand \Eprint [0]{\href }%
\providecommand \doibase [0]{https://doi.org/}%
\providecommand \selectlanguage [0]{\@gobble}%
\providecommand \bibinfo  [0]{\@secondoftwo}%
\providecommand \bibfield  [0]{\@secondoftwo}%
\providecommand \translation [1]{[#1]}%
\providecommand \BibitemOpen [0]{}%
\providecommand \bibitemStop [0]{}%
\providecommand \bibitemNoStop [0]{.\EOS\space}%
\providecommand \EOS [0]{\spacefactor3000\relax}%
\providecommand \BibitemShut  [1]{\csname bibitem#1\endcsname}%
\let\auto@bib@innerbib\@empty
\bibitem [{\citenamefont {Weinberg}(1979)}]{Weinberg:1978kz}%
  \BibitemOpen
  \bibfield  {author} {\bibinfo {author} {\bibfnamefont {S.}~\bibnamefont
  {Weinberg}},\ }\bibfield  {title} {\bibinfo {title} {{Phenomenological
  Lagrangians}},\ }\href {https://doi.org/10.1016/0378-4371(79)90223-1}
  {\bibfield  {journal} {\bibinfo  {journal} {Physica A}\ }\textbf {\bibinfo
  {volume} {96}},\ \bibinfo {pages} {327} (\bibinfo {year} {1979})}\BibitemShut
  {NoStop}%
\bibitem [{\citenamefont {Weinberg}(1990)}]{Weinberg:1990rz}%
  \BibitemOpen
  \bibfield  {author} {\bibinfo {author} {\bibfnamefont {S.}~\bibnamefont
  {Weinberg}},\ }\bibfield  {title} {\bibinfo {title} {{Nuclear forces from
  chiral Lagrangians}},\ }\href {https://doi.org/10.1016/0370-2693(90)90938-3}
  {\bibfield  {journal} {\bibinfo  {journal} {Phys. Lett. B}\ }\textbf
  {\bibinfo {volume} {251}},\ \bibinfo {pages} {288} (\bibinfo {year}
  {1990})}\BibitemShut {NoStop}%
\bibitem [{\citenamefont {Weinberg}(1991)}]{Weinberg:1991um}%
  \BibitemOpen
  \bibfield  {author} {\bibinfo {author} {\bibfnamefont {S.}~\bibnamefont
  {Weinberg}},\ }\bibfield  {title} {\bibinfo {title} {{Effective chiral
  Lagrangians for nucleon - pion interactions and nuclear forces}},\ }\href
  {https://doi.org/10.1016/0550-3213(91)90231-L} {\bibfield  {journal}
  {\bibinfo  {journal} {Nucl. Phys. B}\ }\textbf {\bibinfo {volume} {363}},\
  \bibinfo {pages} {3} (\bibinfo {year} {1991})}\BibitemShut {NoStop}%
\bibitem [{\citenamefont {Machleidt}\ and\ \citenamefont
  {Entem}(2011)}]{Machleidt:2011zz}%
  \BibitemOpen
  \bibfield  {author} {\bibinfo {author} {\bibfnamefont {R.}~\bibnamefont
  {Machleidt}}\ and\ \bibinfo {author} {\bibfnamefont {D.~R.}\ \bibnamefont
  {Entem}},\ }\bibfield  {title} {\bibinfo {title} {{Chiral effective field
  theory and nuclear forces}},\ }\href
  {https://doi.org/10.1016/j.physrep.2011.02.001} {\bibfield  {journal}
  {\bibinfo  {journal} {Phys. Rept.}\ }\textbf {\bibinfo {volume} {503}},\
  \bibinfo {pages} {1} (\bibinfo {year} {2011})},\ \Eprint
  {https://arxiv.org/abs/1105.2919} {arXiv:1105.2919 [nucl-th]} \BibitemShut
  {NoStop}%
\bibitem [{\citenamefont {Epelbaum}\ \emph {et~al.}(2009)\citenamefont
  {Epelbaum}, \citenamefont {Hammer},\ and\ \citenamefont
  {Meissner}}]{Epelbaum:2008ga}%
  \BibitemOpen
  \bibfield  {author} {\bibinfo {author} {\bibfnamefont {E.}~\bibnamefont
  {Epelbaum}}, \bibinfo {author} {\bibfnamefont {H.-W.}\ \bibnamefont
  {Hammer}},\ and\ \bibinfo {author} {\bibfnamefont {U.-G.}\ \bibnamefont
  {Meissner}},\ }\bibfield  {title} {\bibinfo {title} {{Modern Theory of
  Nuclear Forces}},\ }\href {https://doi.org/10.1103/RevModPhys.81.1773}
  {\bibfield  {journal} {\bibinfo  {journal} {Rev. Mod. Phys.}\ }\textbf
  {\bibinfo {volume} {81}},\ \bibinfo {pages} {1773} (\bibinfo {year}
  {2009})},\ \Eprint {https://arxiv.org/abs/0811.1338} {arXiv:0811.1338
  [nucl-th]} \BibitemShut {NoStop}%
\bibitem [{\citenamefont {Hammer}\ \emph {et~al.}(2020)\citenamefont {Hammer},
  \citenamefont {K\"onig},\ and\ \citenamefont {van Kolck}}]{Hammer:2019poc}%
  \BibitemOpen
  \bibfield  {author} {\bibinfo {author} {\bibfnamefont {H.~W.}\ \bibnamefont
  {Hammer}}, \bibinfo {author} {\bibfnamefont {S.}~\bibnamefont {K\"onig}},\
  and\ \bibinfo {author} {\bibfnamefont {U.}~\bibnamefont {van Kolck}},\
  }\bibfield  {title} {\bibinfo {title} {{Nuclear effective field theory:
  status and perspectives}},\ }\href
  {https://doi.org/10.1103/RevModPhys.92.025004} {\bibfield  {journal}
  {\bibinfo  {journal} {Rev. Mod. Phys.}\ }\textbf {\bibinfo {volume} {92}},\
  \bibinfo {pages} {025004} (\bibinfo {year} {2020})},\ \Eprint
  {https://arxiv.org/abs/1906.12122} {arXiv:1906.12122 [nucl-th]} \BibitemShut
  {NoStop}%
\bibitem [{\citenamefont {Melendez}\ \emph {et~al.}(2019)\citenamefont
  {Melendez}, \citenamefont {Furnstahl}, \citenamefont {Phillips},
  \citenamefont {Pratola},\ and\ \citenamefont
  {Wesolowski}}]{Melendez:2019izc}%
  \BibitemOpen
  \bibfield  {author} {\bibinfo {author} {\bibfnamefont {J.~A.}\ \bibnamefont
  {Melendez}}, \bibinfo {author} {\bibfnamefont {R.~J.}\ \bibnamefont
  {Furnstahl}}, \bibinfo {author} {\bibfnamefont {D.~R.}\ \bibnamefont
  {Phillips}}, \bibinfo {author} {\bibfnamefont {M.~T.}\ \bibnamefont
  {Pratola}},\ and\ \bibinfo {author} {\bibfnamefont {S.}~\bibnamefont
  {Wesolowski}},\ }\bibfield  {title} {\bibinfo {title} {{Quantifying
  Correlated Truncation Errors in Effective Field Theory}},\ }\href
  {https://doi.org/10.1103/PhysRevC.100.044001} {\bibfield  {journal} {\bibinfo
   {journal} {Phys. Rev. C}\ }\textbf {\bibinfo {volume} {100}},\ \bibinfo
  {pages} {044001} (\bibinfo {year} {2019})},\ \Eprint
  {https://arxiv.org/abs/1904.10581} {arXiv:1904.10581 [nucl-th]} \BibitemShut
  {NoStop}%
\bibitem [{\citenamefont {Millican}\ \emph {et~al.}(2024)\citenamefont
  {Millican}, \citenamefont {Furnstahl}, \citenamefont {Melendez},
  \citenamefont {Phillips},\ and\ \citenamefont {Pratola}}]{Millican:2024yuz}%
  \BibitemOpen
  \bibfield  {author} {\bibinfo {author} {\bibfnamefont {P.~J.}\ \bibnamefont
  {Millican}}, \bibinfo {author} {\bibfnamefont {R.~J.}\ \bibnamefont
  {Furnstahl}}, \bibinfo {author} {\bibfnamefont {J.~A.}\ \bibnamefont
  {Melendez}}, \bibinfo {author} {\bibfnamefont {D.~R.}\ \bibnamefont
  {Phillips}},\ and\ \bibinfo {author} {\bibfnamefont {M.~T.}\ \bibnamefont
  {Pratola}},\ }\bibfield  {title} {\bibinfo {title} {{Assessing correlated
  truncation errors in modern nucleon-nucleon potentials}},\ }\href
  {https://doi.org/10.1103/PhysRevC.110.044002} {\bibfield  {journal} {\bibinfo
   {journal} {Phys. Rev. C}\ }\textbf {\bibinfo {volume} {110}},\ \bibinfo
  {pages} {044002} (\bibinfo {year} {2024})},\ \Eprint
  {https://arxiv.org/abs/2402.13165} {arXiv:2402.13165 [nucl-th]} \BibitemShut
  {NoStop}%
\bibitem [{\citenamefont {Thim}\ \emph {et~al.}(2024)\citenamefont {Thim},
  \citenamefont {Ekstr\"om},\ and\ \citenamefont {Forss\'en}}]{Thim:2024yks}%
  \BibitemOpen
  \bibfield  {author} {\bibinfo {author} {\bibfnamefont {O.}~\bibnamefont
  {Thim}}, \bibinfo {author} {\bibfnamefont {A.}~\bibnamefont {Ekstr\"om}},\
  and\ \bibinfo {author} {\bibfnamefont {C.}~\bibnamefont {Forss\'en}},\
  }\bibfield  {title} {\bibinfo {title} {{Perturbative computations of
  neutron-proton scattering observables using renormalization-group invariant
  chiral effective field theory up to N3LO}},\ }\href
  {https://doi.org/10.1103/PhysRevC.109.064001} {\bibfield  {journal} {\bibinfo
   {journal} {Phys. Rev. C}\ }\textbf {\bibinfo {volume} {109}},\ \bibinfo
  {pages} {064001} (\bibinfo {year} {2024})},\ \Eprint
  {https://arxiv.org/abs/2402.15325} {arXiv:2402.15325 [nucl-th]} \BibitemShut
  {NoStop}%
\bibitem [{\citenamefont {Ekstr\"om}\ \emph {et~al.}(2023)\citenamefont
  {Ekstr\"om}, \citenamefont {Forss\'en}, \citenamefont {Hagen}, \citenamefont
  {Jansen}, \citenamefont {Jiang},\ and\ \citenamefont
  {Papenbrock}}]{Ekstrom:2022yea}%
  \BibitemOpen
  \bibfield  {author} {\bibinfo {author} {\bibfnamefont {A.}~\bibnamefont
  {Ekstr\"om}}, \bibinfo {author} {\bibfnamefont {C.}~\bibnamefont
  {Forss\'en}}, \bibinfo {author} {\bibfnamefont {G.}~\bibnamefont {Hagen}},
  \bibinfo {author} {\bibfnamefont {G.~R.}\ \bibnamefont {Jansen}}, \bibinfo
  {author} {\bibfnamefont {W.}~\bibnamefont {Jiang}},\ and\ \bibinfo {author}
  {\bibfnamefont {T.}~\bibnamefont {Papenbrock}},\ }\bibfield  {title}
  {\bibinfo {title} {{What is ab initio in nuclear theory?}},\ }\href
  {https://doi.org/10.3389/fphy.2023.1129094} {\bibfield  {journal} {\bibinfo
  {journal} {Front. Phys.}\ }\textbf {\bibinfo {volume} {11}},\ \bibinfo
  {pages} {1129094} (\bibinfo {year} {2023})},\ \Eprint
  {https://arxiv.org/abs/2212.11064} {arXiv:2212.11064 [nucl-th]} \BibitemShut
  {NoStop}%
\bibitem [{\citenamefont {Hergert}(2020)}]{Hergert:2020bxy}%
  \BibitemOpen
  \bibfield  {author} {\bibinfo {author} {\bibfnamefont {H.}~\bibnamefont
  {Hergert}},\ }\bibfield  {title} {\bibinfo {title} {{A Guided Tour of $ab$
  $initio$ Nuclear Many-Body Theory}},\ }\href
  {https://doi.org/10.3389/fphy.2020.00379} {\bibfield  {journal} {\bibinfo
  {journal} {Front. in Phys.}\ }\textbf {\bibinfo {volume} {8}},\ \bibinfo
  {pages} {379} (\bibinfo {year} {2020})},\ \Eprint
  {https://arxiv.org/abs/2008.05061} {arXiv:2008.05061 [nucl-th]} \BibitemShut
  {NoStop}%
\bibitem [{\citenamefont {Entem}\ \emph {et~al.}(2015)\citenamefont {Entem},
  \citenamefont {Kaiser}, \citenamefont {Machleidt},\ and\ \citenamefont
  {Nosyk}}]{Entem:2015xwa}%
  \BibitemOpen
  \bibfield  {author} {\bibinfo {author} {\bibfnamefont {D.~R.}\ \bibnamefont
  {Entem}}, \bibinfo {author} {\bibfnamefont {N.}~\bibnamefont {Kaiser}},
  \bibinfo {author} {\bibfnamefont {R.}~\bibnamefont {Machleidt}},\ and\
  \bibinfo {author} {\bibfnamefont {Y.}~\bibnamefont {Nosyk}},\ }\bibfield
  {title} {\bibinfo {title} {{Dominant contributions to the nucleon-nucleon
  interaction at sixth order of chiral perturbation theory}},\ }\href
  {https://doi.org/10.1103/PhysRevC.92.064001} {\bibfield  {journal} {\bibinfo
  {journal} {Phys. Rev. C}\ }\textbf {\bibinfo {volume} {92}},\ \bibinfo
  {pages} {064001} (\bibinfo {year} {2015})},\ \Eprint
  {https://arxiv.org/abs/1505.03562} {arXiv:1505.03562 [nucl-th]} \BibitemShut
  {NoStop}%
\bibitem [{\citenamefont {Reinert}\ \emph {et~al.}(2018)\citenamefont
  {Reinert}, \citenamefont {Krebs},\ and\ \citenamefont
  {Epelbaum}}]{Reinert:2017usi}%
  \BibitemOpen
  \bibfield  {author} {\bibinfo {author} {\bibfnamefont {P.}~\bibnamefont
  {Reinert}}, \bibinfo {author} {\bibfnamefont {H.}~\bibnamefont {Krebs}},\
  and\ \bibinfo {author} {\bibfnamefont {E.}~\bibnamefont {Epelbaum}},\
  }\bibfield  {title} {\bibinfo {title} {{Semilocal momentum-space regularized
  chiral two-nucleon potentials up to fifth order}},\ }\href
  {https://doi.org/10.1140/epja/i2018-12516-4} {\bibfield  {journal} {\bibinfo
  {journal} {Eur. Phys. J. A}\ }\textbf {\bibinfo {volume} {54}},\ \bibinfo
  {pages} {86} (\bibinfo {year} {2018})},\ \Eprint
  {https://arxiv.org/abs/1711.08821} {arXiv:1711.08821 [nucl-th]} \BibitemShut
  {NoStop}%
\bibitem [{\citenamefont {Nogga}\ \emph {et~al.}(2005)\citenamefont {Nogga},
  \citenamefont {Timmermans},\ and\ \citenamefont {van Kolck}}]{Nogga:2005hy}%
  \BibitemOpen
  \bibfield  {author} {\bibinfo {author} {\bibfnamefont {A.}~\bibnamefont
  {Nogga}}, \bibinfo {author} {\bibfnamefont {R.~G.~E.}\ \bibnamefont
  {Timmermans}},\ and\ \bibinfo {author} {\bibfnamefont {U.}~\bibnamefont {van
  Kolck}},\ }\bibfield  {title} {\bibinfo {title} {{Renormalization of one-pion
  exchange and power counting}},\ }\href
  {https://doi.org/10.1103/PhysRevC.72.054006} {\bibfield  {journal} {\bibinfo
  {journal} {Phys. Rev. C}\ }\textbf {\bibinfo {volume} {72}},\ \bibinfo
  {pages} {054006} (\bibinfo {year} {2005})},\ \Eprint
  {https://arxiv.org/abs/nucl-th/0506005} {arXiv:nucl-th/0506005} \BibitemShut
  {NoStop}%
\bibitem [{\citenamefont {van Kolck}(2020)}]{vanKolck:2020llt}%
  \BibitemOpen
  \bibfield  {author} {\bibinfo {author} {\bibfnamefont {U.}~\bibnamefont {van
  Kolck}},\ }\bibfield  {title} {\bibinfo {title} {{The Problem of
  Renormalization of Chiral Nuclear Forces}},\ }\href
  {https://doi.org/10.3389/fphy.2020.00079} {\bibfield  {journal} {\bibinfo
  {journal} {Front. in Phys.}\ }\textbf {\bibinfo {volume} {8}},\ \bibinfo
  {pages} {79} (\bibinfo {year} {2020})},\ \Eprint
  {https://arxiv.org/abs/2003.06721} {arXiv:2003.06721 [nucl-th]} \BibitemShut
  {NoStop}%
\bibitem [{\citenamefont {Long}\ and\ \citenamefont {van
  Kolck}(2008)}]{Long:2007vp}%
  \BibitemOpen
  \bibfield  {author} {\bibinfo {author} {\bibfnamefont {B.}~\bibnamefont
  {Long}}\ and\ \bibinfo {author} {\bibfnamefont {U.}~\bibnamefont {van
  Kolck}},\ }\bibfield  {title} {\bibinfo {title} {{Renormalization of Singular
  Potentials and Power Counting}},\ }\href
  {https://doi.org/10.1016/j.aop.2008.01.003} {\bibfield  {journal} {\bibinfo
  {journal} {Annals Phys.}\ }\textbf {\bibinfo {volume} {323}},\ \bibinfo
  {pages} {1304} (\bibinfo {year} {2008})},\ \Eprint
  {https://arxiv.org/abs/0707.4325} {arXiv:0707.4325 [quant-ph]} \BibitemShut
  {NoStop}%
\bibitem [{\citenamefont {Pavon~Valderrama}\ and\ \citenamefont
  {Ruiz~Arriola}(2006)}]{PavonValderrama:2005uj}%
  \BibitemOpen
  \bibfield  {author} {\bibinfo {author} {\bibfnamefont {M.}~\bibnamefont
  {Pavon~Valderrama}}\ and\ \bibinfo {author} {\bibfnamefont {E.}~\bibnamefont
  {Ruiz~Arriola}},\ }\bibfield  {title} {\bibinfo {title} {{Renormalization of
  NN interaction with chiral two pion exchange potential: Non-central
  phases}},\ }\href {https://doi.org/10.1103/PhysRevC.74.064004} {\bibfield
  {journal} {\bibinfo  {journal} {Phys. Rev. C}\ }\textbf {\bibinfo {volume}
  {74}},\ \bibinfo {pages} {064004} (\bibinfo {year} {2006})},\ \bibinfo {note}
  {[Erratum: Phys.Rev.C 75, 059905 (2007)]},\ \Eprint
  {https://arxiv.org/abs/nucl-th/0507075} {arXiv:nucl-th/0507075} \BibitemShut
  {NoStop}%
\bibitem [{\citenamefont {Pavon~Valderrama}\ and\ \citenamefont
  {Ruiz~Arriola}(2005)}]{PavonValderrama:2005gu}%
  \BibitemOpen
  \bibfield  {author} {\bibinfo {author} {\bibfnamefont {M.}~\bibnamefont
  {Pavon~Valderrama}}\ and\ \bibinfo {author} {\bibfnamefont {E.}~\bibnamefont
  {Ruiz~Arriola}},\ }\bibfield  {title} {\bibinfo {title} {{Renormalization of
  the deuteron with one pion exchange}},\ }\href
  {https://doi.org/10.1103/PhysRevC.72.054002} {\bibfield  {journal} {\bibinfo
  {journal} {Phys. Rev. C}\ }\textbf {\bibinfo {volume} {72}},\ \bibinfo
  {pages} {054002} (\bibinfo {year} {2005})},\ \Eprint
  {https://arxiv.org/abs/nucl-th/0504067} {arXiv:nucl-th/0504067} \BibitemShut
  {NoStop}%
\bibitem [{\citenamefont {Pavon~Valderrama}(2011)}]{PavonValderrama:2011fcz}%
  \BibitemOpen
  \bibfield  {author} {\bibinfo {author} {\bibfnamefont {M.}~\bibnamefont
  {Pavon~Valderrama}},\ }\bibfield  {title} {\bibinfo {title} {{Perturbative
  Renormalizability of Chiral Two Pion Exchange in Nucleon-Nucleon Scattering:
  P- and D-waves}},\ }\href {https://doi.org/10.1103/PhysRevC.84.064002}
  {\bibfield  {journal} {\bibinfo  {journal} {Phys. Rev. C}\ }\textbf {\bibinfo
  {volume} {84}},\ \bibinfo {pages} {064002} (\bibinfo {year} {2011})},\
  \Eprint {https://arxiv.org/abs/1108.0872} {arXiv:1108.0872 [nucl-th]}
  \BibitemShut {NoStop}%
\bibitem [{\citenamefont {Long}(2013)}]{Long:2013cya}%
  \BibitemOpen
  \bibfield  {author} {\bibinfo {author} {\bibfnamefont {B.}~\bibnamefont
  {Long}},\ }\bibfield  {title} {\bibinfo {title} {{Improved convergence of
  chiral effective field theory for 1S0 of NN scattering}},\ }\href
  {https://doi.org/10.1103/PhysRevC.88.014002} {\bibfield  {journal} {\bibinfo
  {journal} {Phys. Rev. C}\ }\textbf {\bibinfo {volume} {88}},\ \bibinfo
  {pages} {014002} (\bibinfo {year} {2013})},\ \Eprint
  {https://arxiv.org/abs/1304.7382} {arXiv:1304.7382 [nucl-th]} \BibitemShut
  {NoStop}%
\bibitem [{\citenamefont {Pav\'on~Valderrama}\ \emph
  {et~al.}(2017)\citenamefont {Pav\'on~Valderrama}, \citenamefont
  {S\'anchez~S\'anchez}, \citenamefont {Yang}, \citenamefont {Long},
  \citenamefont {Carbonell},\ and\ \citenamefont {van
  Kolck}}]{PavonValderrama:2016lqn}%
  \BibitemOpen
  \bibfield  {author} {\bibinfo {author} {\bibfnamefont {M.}~\bibnamefont
  {Pav\'on~Valderrama}}, \bibinfo {author} {\bibfnamefont {M.}~\bibnamefont
  {S\'anchez~S\'anchez}}, \bibinfo {author} {\bibfnamefont {C.~J.}\
  \bibnamefont {Yang}}, \bibinfo {author} {\bibfnamefont {B.}~\bibnamefont
  {Long}}, \bibinfo {author} {\bibfnamefont {J.}~\bibnamefont {Carbonell}},\
  and\ \bibinfo {author} {\bibfnamefont {U.}~\bibnamefont {van Kolck}},\
  }\bibfield  {title} {\bibinfo {title} {{Power Counting in Peripheral Partial
  Waves: The Singlet Channels}},\ }\href
  {https://doi.org/10.1103/PhysRevC.95.054001} {\bibfield  {journal} {\bibinfo
  {journal} {Phys. Rev. C}\ }\textbf {\bibinfo {volume} {95}},\ \bibinfo
  {pages} {054001} (\bibinfo {year} {2017})},\ \Eprint
  {https://arxiv.org/abs/1611.10175} {arXiv:1611.10175 [nucl-th]} \BibitemShut
  {NoStop}%
\bibitem [{\citenamefont {Valderrama}(2011)}]{Valderrama:2009ei}%
  \BibitemOpen
  \bibfield  {author} {\bibinfo {author} {\bibfnamefont {M.~P.}\ \bibnamefont
  {Valderrama}},\ }\bibfield  {title} {\bibinfo {title} {{Perturbative
  renormalizability of chiral two pion exchange in nucleon-nucleon
  scattering}},\ }\href {https://doi.org/10.1103/PhysRevC.83.024003} {\bibfield
   {journal} {\bibinfo  {journal} {Phys. Rev. C}\ }\textbf {\bibinfo {volume}
  {83}},\ \bibinfo {pages} {024003} (\bibinfo {year} {2011})},\ \Eprint
  {https://arxiv.org/abs/0912.0699} {arXiv:0912.0699 [nucl-th]} \BibitemShut
  {NoStop}%
\bibitem [{\citenamefont {Birse}(2006)}]{Birse:2005um}%
  \BibitemOpen
  \bibfield  {author} {\bibinfo {author} {\bibfnamefont {M.~C.}\ \bibnamefont
  {Birse}},\ }\bibfield  {title} {\bibinfo {title} {{Power counting with
  one-pion exchange}},\ }\href {https://doi.org/10.1103/PhysRevC.74.014003}
  {\bibfield  {journal} {\bibinfo  {journal} {Phys. Rev. C}\ }\textbf {\bibinfo
  {volume} {74}},\ \bibinfo {pages} {014003} (\bibinfo {year} {2006})},\
  \Eprint {https://arxiv.org/abs/nucl-th/0507077} {arXiv:nucl-th/0507077}
  \BibitemShut {NoStop}%
\bibitem [{\citenamefont {Long}\ and\ \citenamefont
  {Yang}(2012{\natexlab{a}})}]{Long:2012ve}%
  \BibitemOpen
  \bibfield  {author} {\bibinfo {author} {\bibfnamefont {B.}~\bibnamefont
  {Long}}\ and\ \bibinfo {author} {\bibfnamefont {C.~J.}\ \bibnamefont
  {Yang}},\ }\bibfield  {title} {\bibinfo {title} {{Short-range nuclear forces
  in singlet channels}},\ }\href {https://doi.org/10.1103/PhysRevC.86.024001}
  {\bibfield  {journal} {\bibinfo  {journal} {Phys. Rev. C}\ }\textbf {\bibinfo
  {volume} {86}},\ \bibinfo {pages} {024001} (\bibinfo {year}
  {2012}{\natexlab{a}})},\ \Eprint {https://arxiv.org/abs/1202.4053}
  {arXiv:1202.4053 [nucl-th]} \BibitemShut {NoStop}%
\bibitem [{\citenamefont {S\'anchez~S\'anchez}\ \emph
  {et~al.}(2018)\citenamefont {S\'anchez~S\'anchez}, \citenamefont {Yang},
  \citenamefont {Long},\ and\ \citenamefont {van
  Kolck}}]{SanchezSanchez:2017tws}%
  \BibitemOpen
  \bibfield  {author} {\bibinfo {author} {\bibfnamefont {M.}~\bibnamefont
  {S\'anchez~S\'anchez}}, \bibinfo {author} {\bibfnamefont {C.~J.}\
  \bibnamefont {Yang}}, \bibinfo {author} {\bibfnamefont {B.}~\bibnamefont
  {Long}},\ and\ \bibinfo {author} {\bibfnamefont {U.}~\bibnamefont {van
  Kolck}},\ }\bibfield  {title} {\bibinfo {title} {{Two-nucleon $^1S_0$
  amplitude zero in chiral effective field theory}},\ }\href
  {https://doi.org/10.1103/PhysRevC.97.024001} {\bibfield  {journal} {\bibinfo
  {journal} {Phys. Rev. C}\ }\textbf {\bibinfo {volume} {97}},\ \bibinfo
  {pages} {024001} (\bibinfo {year} {2018})},\ \Eprint
  {https://arxiv.org/abs/1704.08524} {arXiv:1704.08524 [nucl-th]} \BibitemShut
  {NoStop}%
\bibitem [{\citenamefont {Long}\ and\ \citenamefont
  {Yang}(2012{\natexlab{b}})}]{PhysRevC.85.034002}%
  \BibitemOpen
  \bibfield  {author} {\bibinfo {author} {\bibfnamefont {B.}~\bibnamefont
  {Long}}\ and\ \bibinfo {author} {\bibfnamefont {C.-J.}\ \bibnamefont
  {Yang}},\ }\bibfield  {title} {\bibinfo {title} {Renormalizing chiral nuclear
  forces: Triplet channels},\ }\href
  {https://doi.org/10.1103/PhysRevC.85.034002} {\bibfield  {journal} {\bibinfo
  {journal} {Phys. Rev. C}\ }\textbf {\bibinfo {volume} {85}},\ \bibinfo
  {pages} {034002} (\bibinfo {year} {2012}{\natexlab{b}})}\BibitemShut
  {NoStop}%
\bibitem [{\citenamefont {Yang}(2016)}]{Yang:2016brl}%
  \BibitemOpen
  \bibfield  {author} {\bibinfo {author} {\bibfnamefont {C.~J.}\ \bibnamefont
  {Yang}},\ }\bibfield  {title} {\bibinfo {title} {{Chiral potential
  renormalized in harmonic-oscillator space}},\ }\href
  {https://doi.org/10.1103/PhysRevC.94.064004} {\bibfield  {journal} {\bibinfo
  {journal} {Phys. Rev. C}\ }\textbf {\bibinfo {volume} {94}},\ \bibinfo
  {pages} {064004} (\bibinfo {year} {2016})},\ \Eprint
  {https://arxiv.org/abs/1610.01350} {arXiv:1610.01350 [nucl-th]} \BibitemShut
  {NoStop}%
\bibitem [{\citenamefont {Mishra}\ \emph {et~al.}(2022)\citenamefont {Mishra},
  \citenamefont {Ekstr\"om}, \citenamefont {Hagen}, \citenamefont
  {Papenbrock},\ and\ \citenamefont {Platter}}]{Mishra:2021luw}%
  \BibitemOpen
  \bibfield  {author} {\bibinfo {author} {\bibfnamefont {C.}~\bibnamefont
  {Mishra}}, \bibinfo {author} {\bibfnamefont {A.}~\bibnamefont {Ekstr\"om}},
  \bibinfo {author} {\bibfnamefont {G.}~\bibnamefont {Hagen}}, \bibinfo
  {author} {\bibfnamefont {T.}~\bibnamefont {Papenbrock}},\ and\ \bibinfo
  {author} {\bibfnamefont {L.}~\bibnamefont {Platter}},\ }\bibfield  {title}
  {\bibinfo {title} {{Two-pion exchange as a leading-order contribution in
  chiral effective field theory}},\ }\href
  {https://doi.org/10.1103/PhysRevC.106.024004} {\bibfield  {journal} {\bibinfo
   {journal} {Phys. Rev. C}\ }\textbf {\bibinfo {volume} {106}},\ \bibinfo
  {pages} {024004} (\bibinfo {year} {2022})},\ \Eprint
  {https://arxiv.org/abs/2111.15515} {arXiv:2111.15515 [nucl-th]} \BibitemShut
  {NoStop}%
\bibitem [{\citenamefont {Peng}\ \emph {et~al.}(2022)\citenamefont {Peng},
  \citenamefont {Lyu}, \citenamefont {K\"onig},\ and\ \citenamefont
  {Long}}]{Peng:2021pvo}%
  \BibitemOpen
  \bibfield  {author} {\bibinfo {author} {\bibfnamefont {R.}~\bibnamefont
  {Peng}}, \bibinfo {author} {\bibfnamefont {S.}~\bibnamefont {Lyu}}, \bibinfo
  {author} {\bibfnamefont {S.}~\bibnamefont {K\"onig}},\ and\ \bibinfo {author}
  {\bibfnamefont {B.}~\bibnamefont {Long}},\ }\bibfield  {title} {\bibinfo
  {title} {{Constructing chiral effective field theory around unnatural
  leading-order interactions}},\ }\href
  {https://doi.org/10.1103/PhysRevC.105.054002} {\bibfield  {journal} {\bibinfo
   {journal} {Phys. Rev. C}\ }\textbf {\bibinfo {volume} {105}},\ \bibinfo
  {pages} {054002} (\bibinfo {year} {2022})},\ \Eprint
  {https://arxiv.org/abs/2112.00947} {arXiv:2112.00947 [nucl-th]} \BibitemShut
  {NoStop}%
\bibitem [{\citenamefont {Long}\ and\ \citenamefont
  {Yang}(2011)}]{Long:2011qx}%
  \BibitemOpen
  \bibfield  {author} {\bibinfo {author} {\bibfnamefont {B.}~\bibnamefont
  {Long}}\ and\ \bibinfo {author} {\bibfnamefont {C.~J.}\ \bibnamefont
  {Yang}},\ }\bibfield  {title} {\bibinfo {title} {{Renormalizing chiral
  nuclear forces: a case study of 3P0}},\ }\href
  {https://doi.org/10.1103/PhysRevC.84.057001} {\bibfield  {journal} {\bibinfo
  {journal} {Phys. Rev. C}\ }\textbf {\bibinfo {volume} {84}},\ \bibinfo
  {pages} {057001} (\bibinfo {year} {2011})},\ \Eprint
  {https://arxiv.org/abs/1108.0985} {arXiv:1108.0985 [nucl-th]} \BibitemShut
  {NoStop}%
\bibitem [{\citenamefont {Thim}\ \emph {et~al.}(2023)\citenamefont {Thim},
  \citenamefont {May}, \citenamefont {Ekstr\"om},\ and\ \citenamefont
  {Forss\'en}}]{Thim:2023fnl}%
  \BibitemOpen
  \bibfield  {author} {\bibinfo {author} {\bibfnamefont {O.}~\bibnamefont
  {Thim}}, \bibinfo {author} {\bibfnamefont {E.}~\bibnamefont {May}}, \bibinfo
  {author} {\bibfnamefont {A.}~\bibnamefont {Ekstr\"om}},\ and\ \bibinfo
  {author} {\bibfnamefont {C.}~\bibnamefont {Forss\'en}},\ }\bibfield  {title}
  {\bibinfo {title} {{Bayesian analysis of chiral effective field theory at
  leading order in a modified Weinberg power counting approach}},\ }\href
  {https://doi.org/10.1103/PhysRevC.108.054002} {\bibfield  {journal} {\bibinfo
   {journal} {Phys. Rev. C}\ }\textbf {\bibinfo {volume} {108}},\ \bibinfo
  {pages} {054002} (\bibinfo {year} {2023})},\ \Eprint
  {https://arxiv.org/abs/2302.12624} {arXiv:2302.12624 [nucl-th]} \BibitemShut
  {NoStop}%
\bibitem [{\citenamefont {Thim}(2024)}]{Thim:2024jdv}%
  \BibitemOpen
  \bibfield  {author} {\bibinfo {author} {\bibfnamefont {O.}~\bibnamefont
  {Thim}},\ }\bibfield  {title} {\bibinfo {title} {{Low-Energy Theorems for
  Neutron{\textendash}Proton Scattering in $\chi $EFT Using a Perturbative
  Power Counting}},\ }\href {https://doi.org/10.1007/s00601-024-01938-w}
  {\bibfield  {journal} {\bibinfo  {journal} {Few Body Syst.}\ }\textbf
  {\bibinfo {volume} {65}},\ \bibinfo {pages} {69} (\bibinfo {year} {2024})},\
  \Eprint {https://arxiv.org/abs/2403.10292} {arXiv:2403.10292 [nucl-th]}
  \BibitemShut {NoStop}%
\bibitem [{\citenamefont {Song}\ \emph {et~al.}(2017)\citenamefont {Song},
  \citenamefont {Lazauskas},\ and\ \citenamefont {van Kolck}}]{Song:2016ale}%
  \BibitemOpen
  \bibfield  {author} {\bibinfo {author} {\bibfnamefont {Y.-H.}\ \bibnamefont
  {Song}}, \bibinfo {author} {\bibfnamefont {R.}~\bibnamefont {Lazauskas}},\
  and\ \bibinfo {author} {\bibfnamefont {U.}~\bibnamefont {van Kolck}},\
  }\bibfield  {title} {\bibinfo {title} {{Triton binding energy and
  neutron-deuteron scattering up to next-to-leading order in chiral effective
  field theory}},\ }\href {https://doi.org/10.1103/PhysRevC.96.024002}
  {\bibfield  {journal} {\bibinfo  {journal} {Phys. Rev. C}\ }\textbf {\bibinfo
  {volume} {96}},\ \bibinfo {pages} {024002} (\bibinfo {year} {2017})},\
  \bibinfo {note} {[Erratum: Phys.Rev.C 100, 019901 (2019)]},\ \Eprint
  {https://arxiv.org/abs/1612.09090} {arXiv:1612.09090 [nucl-th]} \BibitemShut
  {NoStop}%
\bibitem [{\citenamefont {Yang}\ \emph {et~al.}(2021)\citenamefont {Yang},
  \citenamefont {Ekstr\"om}, \citenamefont {Forss\'en},\ and\ \citenamefont
  {Hagen}}]{Yang:2020pgi}%
  \BibitemOpen
  \bibfield  {author} {\bibinfo {author} {\bibfnamefont {C.~J.}\ \bibnamefont
  {Yang}}, \bibinfo {author} {\bibfnamefont {A.}~\bibnamefont {Ekstr\"om}},
  \bibinfo {author} {\bibfnamefont {C.}~\bibnamefont {Forss\'en}},\ and\
  \bibinfo {author} {\bibfnamefont {G.}~\bibnamefont {Hagen}},\ }\bibfield
  {title} {\bibinfo {title} {{Power counting in chiral effective field theory
  and nuclear binding}},\ }\href {https://doi.org/10.1103/PhysRevC.103.054304}
  {\bibfield  {journal} {\bibinfo  {journal} {Phys. Rev. C}\ }\textbf {\bibinfo
  {volume} {103}},\ \bibinfo {pages} {054304} (\bibinfo {year} {2021})},\
  \Eprint {https://arxiv.org/abs/2011.11584} {arXiv:2011.11584 [nucl-th]}
  \BibitemShut {NoStop}%
\bibitem [{\citenamefont {Yang}\ \emph {et~al.}(2023)\citenamefont {Yang},
  \citenamefont {Ekstr\"om}, \citenamefont {Forss\'en}, \citenamefont {Hagen},
  \citenamefont {Rupak},\ and\ \citenamefont {van Kolck}}]{Yang:2021vxa}%
  \BibitemOpen
  \bibfield  {author} {\bibinfo {author} {\bibfnamefont {C.~J.}\ \bibnamefont
  {Yang}}, \bibinfo {author} {\bibfnamefont {A.}~\bibnamefont {Ekstr\"om}},
  \bibinfo {author} {\bibfnamefont {C.}~\bibnamefont {Forss\'en}}, \bibinfo
  {author} {\bibfnamefont {G.}~\bibnamefont {Hagen}}, \bibinfo {author}
  {\bibfnamefont {G.}~\bibnamefont {Rupak}},\ and\ \bibinfo {author}
  {\bibfnamefont {U.}~\bibnamefont {van Kolck}},\ }\bibfield  {title} {\bibinfo
  {title} {{The importance of few-nucleon forces in chiral effective field
  theory}},\ }\href {https://doi.org/10.1140/epja/s10050-023-01149-7}
  {\bibfield  {journal} {\bibinfo  {journal} {Eur. Phys. J. A}\ }\textbf
  {\bibinfo {volume} {59}},\ \bibinfo {pages} {233} (\bibinfo {year} {2023})},\
  \Eprint {https://arxiv.org/abs/2109.13303} {arXiv:2109.13303 [nucl-th]}
  \BibitemShut {NoStop}%
\bibitem [{\citenamefont {Gasparyan}\ and\ \citenamefont
  {Epelbaum}(2023)}]{Gasparyan:2022isg}%
  \BibitemOpen
  \bibfield  {author} {\bibinfo {author} {\bibfnamefont {A.~M.}\ \bibnamefont
  {Gasparyan}}\ and\ \bibinfo {author} {\bibfnamefont {E.}~\bibnamefont
  {Epelbaum}},\ }\bibfield  {title} {\bibinfo {title}
  {{\textquotedblleft{}Renormalization-group-invariant effective field
  theory\textquotedblright{} for few-nucleon systems is cutoff dependent}},\
  }\href {https://doi.org/10.1103/PhysRevC.107.034001} {\bibfield  {journal}
  {\bibinfo  {journal} {Phys. Rev. C}\ }\textbf {\bibinfo {volume} {107}},\
  \bibinfo {pages} {034001} (\bibinfo {year} {2023})},\ \Eprint
  {https://arxiv.org/abs/2210.16225} {arXiv:2210.16225 [nucl-th]} \BibitemShut
  {NoStop}%
\bibitem [{\citenamefont {Peng}\ \emph {et~al.}(2024)\citenamefont {Peng},
  \citenamefont {Long},\ and\ \citenamefont {Xu}}]{Peng:2024aiz}%
  \BibitemOpen
  \bibfield  {author} {\bibinfo {author} {\bibfnamefont {R.}~\bibnamefont
  {Peng}}, \bibinfo {author} {\bibfnamefont {B.}~\bibnamefont {Long}},\ and\
  \bibinfo {author} {\bibfnamefont {F.-R.}\ \bibnamefont {Xu}},\ }\bibfield
  {title} {\bibinfo {title} {{Contact operators in renormalization of
  attractive singular potentials}},\ }\href
  {https://doi.org/10.1103/PhysRevC.110.054001} {\bibfield  {journal} {\bibinfo
   {journal} {Phys. Rev. C}\ }\textbf {\bibinfo {volume} {110}},\ \bibinfo
  {pages} {054001} (\bibinfo {year} {2024})},\ \Eprint
  {https://arxiv.org/abs/2407.08342} {arXiv:2407.08342 [nucl-th]} \BibitemShut
  {NoStop}%
\bibitem [{\citenamefont {Yang}(2024)}]{Yang:2024yqv}%
  \BibitemOpen
  \bibfield  {author} {\bibinfo {author} {\bibfnamefont {C.~J.}\ \bibnamefont
  {Yang}},\ }\href@noop {} {\bibinfo {title} {{A further study on the
  renormalization group aspect of perturbative corrections}}} (\bibinfo {year}
  {2024}),\ \Eprint {https://arxiv.org/abs/2410.08845} {arXiv:2410.08845
  [nucl-th]} \BibitemShut {NoStop}%
\bibitem [{\citenamefont {Barrett}\ \emph {et~al.}(2013)\citenamefont
  {Barrett}, \citenamefont {Navratil},\ and\ \citenamefont
  {Vary}}]{Barrett:2013nh}%
  \BibitemOpen
  \bibfield  {author} {\bibinfo {author} {\bibfnamefont {B.~R.}\ \bibnamefont
  {Barrett}}, \bibinfo {author} {\bibfnamefont {P.}~\bibnamefont {Navratil}},\
  and\ \bibinfo {author} {\bibfnamefont {J.~P.}\ \bibnamefont {Vary}},\
  }\bibfield  {title} {\bibinfo {title} {{Ab initio no core shell model}},\
  }\href {https://doi.org/10.1016/j.ppnp.2012.10.003} {\bibfield  {journal}
  {\bibinfo  {journal} {Prog. Part. Nucl. Phys.}\ }\textbf {\bibinfo {volume}
  {69}},\ \bibinfo {pages} {131} (\bibinfo {year} {2013})}\BibitemShut
  {NoStop}%
\bibitem [{\citenamefont {Peng}\ \emph {et~al.}(2025)\citenamefont {Peng},
  \citenamefont {Long},\ and\ \citenamefont {Xu}}]{Peng:2025ykg}%
  \BibitemOpen
  \bibfield  {author} {\bibinfo {author} {\bibfnamefont {R.}~\bibnamefont
  {Peng}}, \bibinfo {author} {\bibfnamefont {B.}~\bibnamefont {Long}},\ and\
  \bibinfo {author} {\bibfnamefont {F.-R.}\ \bibnamefont {Xu}},\ }\href@noop {}
  {\bibinfo {title} {{Perturbative renormalization of chiral nuclear forces at
  subleading order in 3S1-3D1 channel}}} (\bibinfo {year} {2025}),\ \Eprint
  {https://arxiv.org/abs/2508.06838} {arXiv:2508.06838 [nucl-th]} \BibitemShut
  {NoStop}%
\bibitem [{\citenamefont {Wu}\ and\ \citenamefont
  {Long}(2019)}]{PhysRevC.99.024003}%
  \BibitemOpen
  \bibfield  {author} {\bibinfo {author} {\bibfnamefont {S.}~\bibnamefont
  {Wu}}\ and\ \bibinfo {author} {\bibfnamefont {B.}~\bibnamefont {Long}},\
  }\bibfield  {title} {\bibinfo {title} {Perturbative $nn$ scattering in chiral
  effective field theory},\ }\href {https://doi.org/10.1103/PhysRevC.99.024003}
  {\bibfield  {journal} {\bibinfo  {journal} {Phys. Rev. C}\ }\textbf {\bibinfo
  {volume} {99}},\ \bibinfo {pages} {024003} (\bibinfo {year}
  {2019})}\BibitemShut {NoStop}%
\bibitem [{\citenamefont {Peng}\ \emph {et~al.}(2020)\citenamefont {Peng},
  \citenamefont {Lyu},\ and\ \citenamefont {Long}}]{Peng:2020nyz}%
  \BibitemOpen
  \bibfield  {author} {\bibinfo {author} {\bibfnamefont {R.}~\bibnamefont
  {Peng}}, \bibinfo {author} {\bibfnamefont {S.}~\bibnamefont {Lyu}},\ and\
  \bibinfo {author} {\bibfnamefont {B.}~\bibnamefont {Long}},\ }\bibfield
  {title} {\bibinfo {title} {{Perturbative chiral nucleon\textendash{}nucleon
  potential for the $^3P_0$ partial wave}},\ }\href
  {https://doi.org/10.1088/1572-9494/aba251} {\bibfield  {journal} {\bibinfo
  {journal} {Commun. Theor. Phys.}\ }\textbf {\bibinfo {volume} {72}},\
  \bibinfo {pages} {095301} (\bibinfo {year} {2020})},\ \Eprint
  {https://arxiv.org/abs/2011.13186} {arXiv:2011.13186 [nucl-th]} \BibitemShut
  {NoStop}%
\bibitem [{\citenamefont {Stoks}\ \emph {et~al.}(1993)\citenamefont {Stoks},
  \citenamefont {Klomp}, \citenamefont {Rentmeester},\ and\ \citenamefont
  {de~Swart}}]{Stoks:1993tb}%
  \BibitemOpen
  \bibfield  {author} {\bibinfo {author} {\bibfnamefont {V.~G.~J.}\
  \bibnamefont {Stoks}}, \bibinfo {author} {\bibfnamefont {R.~A.~M.}\
  \bibnamefont {Klomp}}, \bibinfo {author} {\bibfnamefont {M.~C.~M.}\
  \bibnamefont {Rentmeester}},\ and\ \bibinfo {author} {\bibfnamefont {J.~J.}\
  \bibnamefont {de~Swart}},\ }\bibfield  {title} {\bibinfo {title} {{Partial
  wave analaysis of all nucleon-nucleon scattering data below 350-MeV}},\
  }\href {https://doi.org/10.1103/PhysRevC.48.792} {\bibfield  {journal}
  {\bibinfo  {journal} {Phys. Rev. C}\ }\textbf {\bibinfo {volume} {48}},\
  \bibinfo {pages} {792} (\bibinfo {year} {1993})}\BibitemShut {NoStop}%
\bibitem [{\citenamefont {Shi}\ \emph {et~al.}(2022)\citenamefont {Shi},
  \citenamefont {Peng}, \citenamefont {Liu}, \citenamefont {Lyu},\ and\
  \citenamefont {Long}}]{Shi:2022blm}%
  \BibitemOpen
  \bibfield  {author} {\bibinfo {author} {\bibfnamefont {W.}~\bibnamefont
  {Shi}}, \bibinfo {author} {\bibfnamefont {R.}~\bibnamefont {Peng}}, \bibinfo
  {author} {\bibfnamefont {T.-X.}\ \bibnamefont {Liu}}, \bibinfo {author}
  {\bibfnamefont {S.}~\bibnamefont {Lyu}},\ and\ \bibinfo {author}
  {\bibfnamefont {B.}~\bibnamefont {Long}},\ }\bibfield  {title} {\bibinfo
  {title} {{Perturbative calculations of deuteron form factors}},\ }\href
  {https://doi.org/10.1103/PhysRevC.106.015505} {\bibfield  {journal} {\bibinfo
   {journal} {Phys. Rev. C}\ }\textbf {\bibinfo {volume} {106}},\ \bibinfo
  {pages} {015505} (\bibinfo {year} {2022})},\ \Eprint
  {https://arxiv.org/abs/2205.02000} {arXiv:2205.02000 [nucl-th]} \BibitemShut
  {NoStop}%
\bibitem [{\citenamefont {Navratil}\ \emph {et~al.}(2000)\citenamefont
  {Navratil}, \citenamefont {Kamuntavicius},\ and\ \citenamefont
  {Barrett}}]{Navratil:1999pw}%
  \BibitemOpen
  \bibfield  {author} {\bibinfo {author} {\bibfnamefont {P.}~\bibnamefont
  {Navratil}}, \bibinfo {author} {\bibfnamefont {G.~P.}\ \bibnamefont
  {Kamuntavicius}},\ and\ \bibinfo {author} {\bibfnamefont {B.~R.}\
  \bibnamefont {Barrett}},\ }\bibfield  {title} {\bibinfo {title} {{Few nucleon
  systems in translationally invariant harmonic oscillator basis}},\ }\href
  {https://doi.org/10.1103/PhysRevC.61.044001} {\bibfield  {journal} {\bibinfo
  {journal} {Phys. Rev. C}\ }\textbf {\bibinfo {volume} {61}},\ \bibinfo
  {pages} {044001} (\bibinfo {year} {2000})},\ \Eprint
  {https://arxiv.org/abs/nucl-th/9907054} {arXiv:nucl-th/9907054} \BibitemShut
  {NoStop}%
\bibitem [{\citenamefont {Thim}(2025)}]{py-ncsm}%
  \BibitemOpen
  \bibfield  {author} {\bibinfo {author} {\bibfnamefont {O.}~\bibnamefont
  {Thim}},\ }\href@noop {} {\bibinfo {title} {{py-ncsm}: Python code for the
  no-core shell model}},\ \bibinfo {howpublished}
  {\url{https://github.com/othim/py-ncsm}} (\bibinfo {year} {2025}),\ \bibinfo
  {note} {github repository}\BibitemShut {NoStop}%
\bibitem [{\citenamefont {Kamuntavicius}\ \emph {et~al.}(1999)\citenamefont
  {Kamuntavicius}, \citenamefont {Navratil}, \citenamefont {Barrett},
  \citenamefont {Sapragonaite},\ and\ \citenamefont
  {Kalinauskas}}]{Kamuntavicius:1999eu}%
  \BibitemOpen
  \bibfield  {author} {\bibinfo {author} {\bibfnamefont {G.~P.}\ \bibnamefont
  {Kamuntavicius}}, \bibinfo {author} {\bibfnamefont {P.}~\bibnamefont
  {Navratil}}, \bibinfo {author} {\bibfnamefont {B.~R.}\ \bibnamefont
  {Barrett}}, \bibinfo {author} {\bibfnamefont {G.}~\bibnamefont
  {Sapragonaite}},\ and\ \bibinfo {author} {\bibfnamefont {R.~K.}\ \bibnamefont
  {Kalinauskas}},\ }\bibfield  {title} {\bibinfo {title} {{Isoscalar
  Hamiltonians for light atomic nuclei}},\ }\href
  {https://doi.org/10.1103/PhysRevC.60.044304} {\bibfield  {journal} {\bibinfo
  {journal} {Phys. Rev. C}\ }\textbf {\bibinfo {volume} {60}},\ \bibinfo
  {pages} {044304} (\bibinfo {year} {1999})},\ \Eprint
  {https://arxiv.org/abs/nucl-th/9907047} {arXiv:nucl-th/9907047} \BibitemShut
  {NoStop}%
\bibitem [{\citenamefont {Forss\'en}\ \emph {et~al.}(2018)\citenamefont
  {Forss\'en}, \citenamefont {Carlsson}, \citenamefont {Johansson},
  \citenamefont {S\"a\"af}, \citenamefont {Bansal}, \citenamefont {Hagen},\
  and\ \citenamefont {Papenbrock}}]{Forssen:2017wei}%
  \BibitemOpen
  \bibfield  {author} {\bibinfo {author} {\bibfnamefont {C.}~\bibnamefont
  {Forss\'en}}, \bibinfo {author} {\bibfnamefont {B.~D.}\ \bibnamefont
  {Carlsson}}, \bibinfo {author} {\bibfnamefont {H.~T.}\ \bibnamefont
  {Johansson}}, \bibinfo {author} {\bibfnamefont {D.}~\bibnamefont {S\"a\"af}},
  \bibinfo {author} {\bibfnamefont {A.}~\bibnamefont {Bansal}}, \bibinfo
  {author} {\bibfnamefont {G.}~\bibnamefont {Hagen}},\ and\ \bibinfo {author}
  {\bibfnamefont {T.}~\bibnamefont {Papenbrock}},\ }\bibfield  {title}
  {\bibinfo {title} {{Large-scale exact diagonalizations reveal low-momentum
  scales of nuclei}},\ }\href {https://doi.org/10.1103/PhysRevC.97.034328}
  {\bibfield  {journal} {\bibinfo  {journal} {Phys. Rev. C}\ }\textbf {\bibinfo
  {volume} {97}},\ \bibinfo {pages} {034328} (\bibinfo {year} {2018})},\
  \Eprint {https://arxiv.org/abs/1712.09951} {arXiv:1712.09951 [nucl-th]}
  \BibitemShut {NoStop}%
\bibitem [{\citenamefont {Wendt}\ \emph {et~al.}(2015)\citenamefont {Wendt},
  \citenamefont {Forss\'en}, \citenamefont {Papenbrock},\ and\ \citenamefont
  {S\"a\"af}}]{Wendt:2015nba}%
  \BibitemOpen
  \bibfield  {author} {\bibinfo {author} {\bibfnamefont {K.~A.}\ \bibnamefont
  {Wendt}}, \bibinfo {author} {\bibfnamefont {C.}~\bibnamefont {Forss\'en}},
  \bibinfo {author} {\bibfnamefont {T.}~\bibnamefont {Papenbrock}},\ and\
  \bibinfo {author} {\bibfnamefont {D.}~\bibnamefont {S\"a\"af}},\ }\bibfield
  {title} {\bibinfo {title} {{Infrared length scale and extrapolations for the
  no-core shell model}},\ }\href {https://doi.org/10.1103/PhysRevC.91.061301}
  {\bibfield  {journal} {\bibinfo  {journal} {Phys. Rev. C}\ }\textbf {\bibinfo
  {volume} {91}},\ \bibinfo {pages} {061301} (\bibinfo {year} {2015})},\
  \Eprint {https://arxiv.org/abs/1503.07144} {arXiv:1503.07144 [nucl-th]}
  \BibitemShut {NoStop}%
\bibitem [{\citenamefont {Kukulin}\ and\ \citenamefont
  {Pomerantsev}(1978)}]{Kukulin:1978he}%
  \BibitemOpen
  \bibfield  {author} {\bibinfo {author} {\bibfnamefont {V.~I.}\ \bibnamefont
  {Kukulin}}\ and\ \bibinfo {author} {\bibfnamefont {V.~N.}\ \bibnamefont
  {Pomerantsev}},\ }\bibfield  {title} {\bibinfo {title} {{The Orthogonal
  Projection Method in Scattering Theory}},\ }\href
  {https://doi.org/10.1016/0003-4916(78)90069-6} {\bibfield  {journal}
  {\bibinfo  {journal} {Annals Phys.}\ }\textbf {\bibinfo {volume} {111}},\
  \bibinfo {pages} {330} (\bibinfo {year} {1978})}\BibitemShut {NoStop}%
\bibitem [{\citenamefont {Purcell}\ \emph {et~al.}(2010)\citenamefont
  {Purcell}, \citenamefont {Kelley}, \citenamefont {Kwan}, \citenamefont
  {Sheu},\ and\ \citenamefont {Weller}}]{Purcell:2010hka}%
  \BibitemOpen
  \bibfield  {author} {\bibinfo {author} {\bibfnamefont {J.~E.}\ \bibnamefont
  {Purcell}}, \bibinfo {author} {\bibfnamefont {J.~H.}\ \bibnamefont {Kelley}},
  \bibinfo {author} {\bibfnamefont {E.}~\bibnamefont {Kwan}}, \bibinfo {author}
  {\bibfnamefont {C.~G.}\ \bibnamefont {Sheu}},\ and\ \bibinfo {author}
  {\bibfnamefont {H.~R.}\ \bibnamefont {Weller}},\ }\bibfield  {title}
  {\bibinfo {title} {{Energy levels of light nuclei A = 3}},\ }\href
  {https://doi.org/10.1016/j.nuclphysa.2010.08.012} {\bibfield  {journal}
  {\bibinfo  {journal} {Nucl. Phys. A}\ }\textbf {\bibinfo {volume} {848}},\
  \bibinfo {pages} {1} (\bibinfo {year} {2010})}\BibitemShut {NoStop}%
\bibitem [{\citenamefont {Workman}\ \emph {et~al.}()\citenamefont {Workman}
  \emph {et~al.}}]{PDG_2022}%
  \BibitemOpen
  \bibfield  {author} {\bibinfo {author} {\bibfnamefont {R.}~\bibnamefont
  {Workman}} \emph {et~al.} (\bibinfo {collaboration} {Particle Data Group}),\
  }\href@noop {} {\bibinfo {title} {Review of particle physics}},\ \bibinfo
  {note} {to be published (2022)}\BibitemShut {NoStop}%
\bibitem [{\citenamefont {Pavon~Valderrama}(2025)}]{PavonValderrama:2025zzk}%
  \BibitemOpen
  \bibfield  {author} {\bibinfo {author} {\bibfnamefont {M.}~\bibnamefont
  {Pavon~Valderrama}},\ }\href@noop {} {\bibinfo {title} {{Regulator
  constraints for the perturbative renormalizability of attractive triplets}}}
  (\bibinfo {year} {2025}),\ \Eprint {https://arxiv.org/abs/2509.23855}
  {arXiv:2509.23855 [nucl-th]} \BibitemShut {NoStop}%
\bibitem [{\citenamefont {Weinberg}(1992)}]{Weinberg:1992yk}%
  \BibitemOpen
  \bibfield  {author} {\bibinfo {author} {\bibfnamefont {S.}~\bibnamefont
  {Weinberg}},\ }\bibfield  {title} {\bibinfo {title} {{Three body interactions
  among nucleons and pions}},\ }\href
  {https://doi.org/10.1016/0370-2693(92)90099-P} {\bibfield  {journal}
  {\bibinfo  {journal} {Phys. Lett. B}\ }\textbf {\bibinfo {volume} {295}},\
  \bibinfo {pages} {114} (\bibinfo {year} {1992})},\ \Eprint
  {https://arxiv.org/abs/hep-ph/9209257} {arXiv:hep-ph/9209257} \BibitemShut
  {NoStop}%
\bibitem [{\citenamefont {van Kolck}(1994)}]{vanKolck:1994yi}%
  \BibitemOpen
  \bibfield  {author} {\bibinfo {author} {\bibfnamefont {U.}~\bibnamefont {van
  Kolck}},\ }\bibfield  {title} {\bibinfo {title} {{Few nucleon forces from
  chiral Lagrangians}},\ }\href {https://doi.org/10.1103/PhysRevC.49.2932}
  {\bibfield  {journal} {\bibinfo  {journal} {Phys. Rev. C}\ }\textbf {\bibinfo
  {volume} {49}},\ \bibinfo {pages} {2932} (\bibinfo {year}
  {1994})}\BibitemShut {NoStop}%
\bibitem [{\citenamefont {Stapp}\ \emph {et~al.}(1957)\citenamefont {Stapp},
  \citenamefont {Ypsilantis},\ and\ \citenamefont {Metropolis}}]{Stapp:1956mz}%
  \BibitemOpen
  \bibfield  {author} {\bibinfo {author} {\bibfnamefont {H.~P.}\ \bibnamefont
  {Stapp}}, \bibinfo {author} {\bibfnamefont {T.~J.}\ \bibnamefont
  {Ypsilantis}},\ and\ \bibinfo {author} {\bibfnamefont {N.}~\bibnamefont
  {Metropolis}},\ }\bibfield  {title} {\bibinfo {title} {{Phase shift analysis
  of 310-MeV proton proton scattering experiments}},\ }\href
  {https://doi.org/10.1103/PhysRev.105.302} {\bibfield  {journal} {\bibinfo
  {journal} {Phys. Rev.}\ }\textbf {\bibinfo {volume} {105}},\ \bibinfo {pages}
  {302} (\bibinfo {year} {1957})}\BibitemShut {NoStop}%
\end{thebibliography}%

\clearpage\newpage
\appendix
\begin{widetext}
\section{Toy problem of exceptional behavior in the non-singular $^1S_0$ channel \label{app:1S0}}

The appearance of exceptional points is connected to the properties of the LO wave function, i.e., the distorted wave when computing subleading corrections. It is not a feature intrinsic to singular potentials, but can also appear as a non-singular potential is made more attractive. We will illustrate this, and the role of the LO wave function, with a toy scenario that involves variation of the axial coupling, $g_A$, in the \LO{} potential.

We consider the $^1S_0$ partial wave and the LO potential according to \cref{tab:potentials_PC}, with $\Lambda=500$~MeV. We vary $g_A$, which controls the strength of OPE, and adjust the LO LEC to reproduce the empirical $^1S_0$ phase shift \cite{Stoks:1993tb} at $\Tl=40$~MeV. The resulting value of the LEC, $C^{(0)}_{^1S_0}$, is shown in the left panel of \cref{fig:LECs_1S0} and a divergence is observed, analogous to the limit-cycle-like divergence at LO in \cref{fig:LECs_3P0,fig:LECs_3S-D1}.

Furthermore, we add the \NLO{} correction in $^1S_0$ perturbatively (see \cref{tab:potentials_PC}) and the two LECs $C^{(1)}_{^1S_0}$ and $D^{(0)}_{^1S_0}$ are fixed by reproducing the empirical phase shifts at $\Tl=50$~MeV and $\Tl=80$~MeV. The \NLO{} LECs as a function of $g_A$ are shown in the middle and right panels of \cref{fig:LECs_1S0}. Here we observe two divergences, one where the LO LEC diverges and one for a slightly lower value of $g_A$. We can now see that \cref{fig:LECs_1S0} is completely analogous to \cref{fig:LECs_3P0}, but now $g_A$ is varied instead of $\Lambda$.

The predicted $^1S_0$ phase shift at $\Tl=60$~MeV is shown in \cref{fig:1S0_phases}. It can be observed that this prediction diverges at $g_A \approx 4.023$, which is the same value indicated by the red vertical line in \cref{fig:LECs_1S0}. This means that for $g_A\approx 4.023$ we cannot find NLO LECs with the renormalization conditions used. The situation is completely analogous to what happens at an exceptional cutoff.

This toy computation illustrates that the LO wave function seems to behave similarly when the potential $(i)$ is singular and the cutoff is increased, and $(ii)$ is non-singular and its strength is increased. This can explain the qualitative similarity of the \cref{fig:LECs_3P0,fig:LECs_1S0}, even though the independent variable is different. Therefore, it might be beneficial to study the problem of exceptional cutoffs in the Long and Yang PC (and MWPCs in general) from the perspective of an ``exceptional strength'' of the LO potential, or simply an ``exceptional'' LO wave function. Interestingly, a recent work \cite{PavonValderrama:2025zzk} couples the appearance of exceptional points to the way the potentials are regulated.

\begin{figure*}
    \centering
    \includegraphics[width=\textwidth]{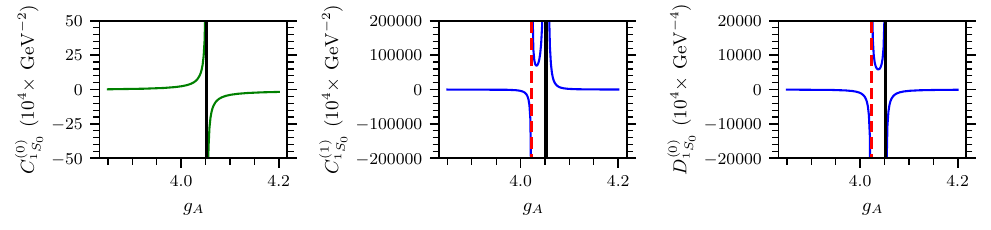}
    \caption{LECs at LO (left panel) and \NLO{} (middle and right panel) in the $^1S_0$ channel as function of $g_A$ for cutoff $\Lambda=500$~MeV. The solid vertical line shows the location where the LO LEC diverges ($g_A \approx 4.052$), while the dashed vertical line shows the additional value of $g_A$ where the \NLO{} LECs diverge ($g_A \approx 4.023$).}
    \label{fig:LECs_1S0}
\end{figure*}

\begin{figure}
    \centering
    \includegraphics[width=0.45\textwidth]{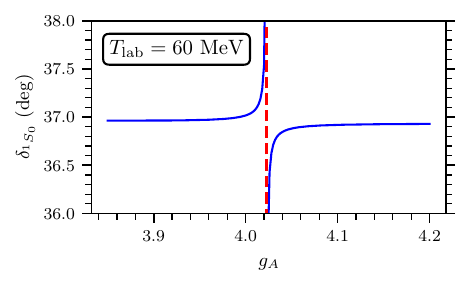}
    \caption{Predicted phase shift at $\Tl=60$~MeV in the $^1S_0$ channel at \NLO{} as a function of $g_A$. Here, $\Lambda=500$~MeV. The red dashed line indicates where the prediction diverges and has the same location as in \cref{fig:LECs_1S0}.}
    \label{fig:1S0_phases}
\end{figure}

\section{Exceptional cutoffs in the $\tS$ channel \label{app:exceptional_cutoffs}}

To study exceptional cutoffs in the $\tS$ channel, we can proceed in the same way as for the $^3P_0$ channel~\cite{Gasparyan:2022isg}. The goal is to arrive at an equation similar to \cref{eq:A_3P0_in_text}, but for the \NNLO{} LECs in the $\tS$ channel.

We consider neutron-proton scattering, where a neutron with laboratory kinetic energy $\Tl$ impinges on a proton. The \LO{} amplitude in the $\tS$ channel, $\TLO$, is obtained (non-perturbatively) by solving the LS-equation 
\begin{equation}
    \TLO = \VLO + \VLO G^+_0\TLO,
\end{equation}
where the free resolvent is given by
\begin{equation}
    G^+_0 = \left(E - H_0 + i\epsilon\right)^{-1},
\end{equation}
and $H_0 = \bm{p}^2/m_N$. Here, $\bm{p}$ denotes the \NN{} relative momentum in the \cm{} frame and $m_N$ denotes the nucleon mass. Explicitly, we solve the partial wave projected LS-equation
\begin{equation}
    T^{(0)}_{\l'\l}(p',p;k) = V^{(0)}_{\l'\l}(p',p) + \sum_{\l''}\int_0^\infty dp'' p''^2 V^{(0)}_{\l'\l''}(p',p'') \frac{m_N}{\pon^2-p''^2 + i \epsilon} T^{(0)}_{\l''\l}(p'',p;k)
    \label{eq:3S-D1_LS}
\end{equation}
to obtain the scattering amplitude $T^{(0)}_{\l'\l}(p',p;k)$. Here, $p\ (p')$ denotes the modulus of the ingoing (outgoing) relative momentum in the $np$ scattering process, and $\l$ ($\l')$ denotes the ongoing (outgoing) angular momentum quantum number. The modulus of the on-shell momentum, $\pon$, is related to the laboratory kinetic energy, $\Tl$, as 
\begin{equation}
    \pon^2 = \frac{m_p^2 T_\mathrm{lab} (2m_n + T_\mathrm{lab})}{(m_n+m_p)^2 + 2m_p T_\mathrm{lab}}.
    \label{eq:q_on_shell}
\end{equation}

The \NLO{} contribution in the $\tS$ channel vanishes while the \NNLO{} potential is non-zero and gives the \NNLO{} amplitude as
\begin{align}
    \TNNLO &= T^{(2)}_\pi + T^{(2)}_\mathrm{ct} \nonumber \\ 
    T^{(2)}_\pi &= \Omega^\dagger_-V^{(2)}_{2\pi} \Omega_+ \label{eq:3S-D1_N2LO}\\
    T^{(2)}_\mathrm{ct} &= \Omega^\dagger_-V^{(2)}_\mathrm{ct} \Omega_+, \nonumber
\end{align}
where we separate the parts coming from the two-pion exchange and the contact terms. The M\o ller wave operators are defined as $\Omega_+ = \mathds{1} + G^+_0\TLO$ and $\Omega^\dagger_- = \mathds{1} + \TLO G^+_0$. 

We can write the \NNLO{} contact potential as a matrix in $\l$ \cite{Thim:2024yks}
\begin{equation}
    V^{(2)}_\mathrm{ct} = \frac{1}{(2\pi)^{3}}\begin{pmatrix} C^{(1)}_{^3S_1} + D^{(0)}_{^3S_1}(p'^2 + p^2) &  D^{(0)}_{SD} p^2\\D^{(0)}_{SD}p'^2 & 0 \end{pmatrix}
\end{equation}
and the contact part of the on-shell $T$-matrix can be decomposed as
\begin{equation}
    T^{(2)}_\mathrm{ct} = \Omega^\dagger_- V^{(2)}_\mathrm{ct}  \Omega_+ = T^{(2)}_0 C^{(1)}_{^3S_1}+ T^{(2)}_1 D^{(0)}_{^3S_1} +  T^{(2)}_2 D^{(0)}_{SD}.
\end{equation}
An explicit calculation gives
\begin{align}
    T^{(2)}_\mathrm{ct}(k,k;k) &= 
    \begin{pmatrix}
        \psi_{00,0}^2 & \psi_{00,0}\psi_{20,0} \\ \psi_{00,0}\psi_{20,0}& \psi^2_{20,0}
    \end{pmatrix} C^{(1)}_{^3S_1} \nonumber \\ &+\begin{pmatrix}
        2\psi_{00,0}\psi_{00,2} & \psi_{00,0}\psi_{20,2} + \psi_{00,2}\psi_{20,0} \\ \psi_{00,0}\psi_{20,2} + \psi_{00,2}\psi_{20,0} & 2\psi_{20,0}(k)\psi_{20,2}
    \end{pmatrix} D^{(0)}_{^3S_1}  \label{eq:T_ct_decomp_3S-D1} \\  &+  \begin{pmatrix}
        2\psi_{02,2}\psi_{00,0} & \psi_{22,2}\psi_{00,0} + \psi_{02,2}\psi_{20,0} \\ \psi_{22,2}\psi_{00,0} + \psi_{02,2}\psi_{20,0} & 2\psi_{22,2}\psi_{20,0}
    \end{pmatrix} D^{(0)}_{SD}
    \nonumber
\end{align}
where
\begin{align}
    (2\pi)^{3/2}\psi_{\l'\l,n}(k) \equiv k^n\exp{-(k/\Lambda)^6} \delta_{\l'\l}+ \int_0^\infty dq \ q^{2+n}  \exp{-(q/\Lambda)^6}G^+_0(q;k)T^{(0)}_{\l'\l}(k,q;k).
\end{align}
Note that the $\Lambda$- and $k$- dependence of $\psi_{\l'\l,n}$ is not explicitly shown.
We use Eq. (C27) in Ref.~\cite{Thim:2024yks} to relate the $T$-matrix with the phase shifts $\delta^{(\nu)}_\l$ and mixing angle $\epsilon^{(\nu)}$ in the $\tS$ channel as
\begin{equation}
       \begin{pmatrix} S_{00}^{(2)}(\pon,\pon) \\ S_{02}^{(2)}(\pon,\pon) \\ S_{22}^{(2)}(\pon,\pon)  \end{pmatrix} = D(\pon) \begin{pmatrix} \epsilon^{(2)}(\pon) \\ \delta^{(2)}_0(\pon) \\ \delta^{(2)}_2(\pon) \end{pmatrix} \\
\end{equation}
where the matrix $D(k)$ depends on the LO amplitude, and we use the notation $S^{(2)}_{\l'\l}(k,k) = -\rho(k) T^{(2)}_{\l'\l}(k,k;k)$. Using \cref{eq:3S-D1_N2LO} and \cref{eq:T_ct_decomp_3S-D1} we obtain the decomposition
\begin{equation}
       \begin{pmatrix} S_{00}^{(2)}(\pon,\pon) \\ S_{02}^{(2)}(\pon,\pon) \\ S_{22}^{(2)}(\pon,\pon)  \end{pmatrix} = B(\pon) \begin{pmatrix} C^{(1)}_{^3S_1} \\D^{(0)}_{^3S_1}\\D^{(0)}_{SD} \end{pmatrix}  -\rho(k)\begin{pmatrix}
           T^{(2)}_{\mathrm{\pi},00}(\pon,\pon;\pon) \\T^{(2)}_{\mathrm{\pi},02}(\pon,\pon;\pon) \\ T^{(2)}_{\mathrm{\pi},22}(\pon,\pon;\pon)
       \end{pmatrix},
\end{equation}
where $B(\pon)$ is constructed using \cref{eq:T_ct_decomp_3S-D1}.

The phase shifts can now be expressed as
\begin{equation}
    \vec{\delta}^{(2)}(\pon) \equiv \begin{pmatrix} \epsilon^{(2)}(k) \\ \delta^{(2)}_0(k) \\ \delta^{(2)}_2(k) \end{pmatrix} = D^{-1}(\pon)B(\pon) \begin{pmatrix} C^{(1)}_{^3S_1} \\D^{(0)}_{^3S_1}\\D^{(0)}_{SD} \end{pmatrix}\underbrace{-\rho(k)D^{-1}(\pon)\begin{pmatrix}
           T^{(2)}_{\mathrm{\pi},00} \\ T^{(2)}_{\mathrm{\pi},02} \\ T^{(2)}_{\mathrm{\pi},22}
       \end{pmatrix}}_{\equiv \vec{\delta}^{(2)}_\pi(\pon)},
\end{equation}
where we define the short-hand notation
\begin{equation}
    \vec{\delta}^{(2)}(\pon) = C(\pon)\vec{\alpha}^{(2)} + \vec{\delta}^{(2)}_{\pi}(\pon),
\end{equation}
with $C\equiv D^{-1}B$, and LECs $\left(\vec{\alpha}^{(2)}\right)^T \equiv (C^{(1)}_{^3S_1},D^{(0)}_{^3S_1},D^{(0)}_{SD})$.

The renormalization conditions used to fix the LECs at \NNLO{} read \cite{Thim:2024yks}
\begin{equation}
    \begin{pmatrix}
        \delta^{(0)}_0(k_1)+\delta^{(2)}_0(k_1) \\ \delta^{(0)}_0(k_2) + \delta^{(2)}_0(k_2) \\ \epsilon^{(0)}(k_2) + \epsilon^{(2)}(k_2)
    \end{pmatrix} = \begin{pmatrix}
       \delta_\mathrm{0,exp}(k_1)\\ \delta_\mathrm{0,exp}(k_2) \\ \epsilon_\mathrm{exp}(k_2) 
    \end{pmatrix}
\end{equation}
where $k_1$ $(k_2)$ correspond to $\Tl=30$~MeV $(\Tl=50$~MeV), and $\delta^{(\nu)}_0$ are the phase shift in the $\l=0$ channel while $\epsilon^{(\nu)}$ is the mixing angle. This gives a matrix equation for the LECs 
\begin{equation}
\begin{pmatrix}
        - & C_{1i}(k_1) & - \\  - & C_{1i}(k_2) & -  \\  - & C_{0i}(k_1) & - 
    \end{pmatrix} \vec{\alpha}^{(2)} + 
    \begin{pmatrix}
        \delta^{(2)}_{0,\pi}(k_1) \\ \delta^{(2)}_{0,\pi}(k_2)\\ \epsilon^{(2)}_\pi(k_2)
    \end{pmatrix}
    = \begin{pmatrix}
        \delta_\mathrm{0,exp}(k_1) - \delta^{(0)}_0(k_1) \\ \delta_\mathrm{0,exp}(k_2) -\delta^{(0)}_0(k_2) \\ \epsilon_\mathrm{exp}(k_2) -\epsilon^{(0)}(k_2) 
    \end{pmatrix}
\end{equation}
which can be written as
\begin{equation}
    A \vec{\alpha}^{(2)} = \begin{pmatrix}
        \delta_\mathrm{0,exp}(k_1) - \delta^{(0)}_0(k_1) -\delta^{(2)}_{0,\pi}(k_1)\\ \delta_\mathrm{0,exp}(k_2) -\delta^{(0)}_0(k_2) - \delta^{(2)}_{0,\pi}(k_2)\\ \epsilon_\mathrm{exp}(k_2) -\epsilon^{(0)}(k_2) -\epsilon^{(2)}_\pi(k_2)
    \end{pmatrix} \equiv \vec{\delta}
    \label{eq:A_3S-D1_app}
\end{equation}
This equation is the equivalent of \cref{eq:A_3P0_in_text}, but now for the $\tS$ channel.

\Cref{fig:LECs_3S-D1_long} shows the LECs at LO and \NNLO{} as a function of the cutoff for a larger cutoff interval than \cref{fig:LECs_3S-D1}. The vertical lines indicate locations of exceptional- and limit-cycle-like cutoffs, whose locations coincide with the zeros of the determinant of $A$ in \cref{eq:A_3S-D1_app}. We note a more irregular appearance of exceptional cutoffs compared to the $^3P_0$ channel. Furthermore, \cref{fig:3S-D1_phases_Lambda} shows predicted phase shifts at $\Tl=100$~MeV and divergences at the exceptional cutoffs can be observed. This is consistent with the observed divergences in the deuteron ground-state energy in \cref{fig:deuteron_Lambda}.

\begin{figure*}
    \centering
    \includegraphics[width=\textwidth]{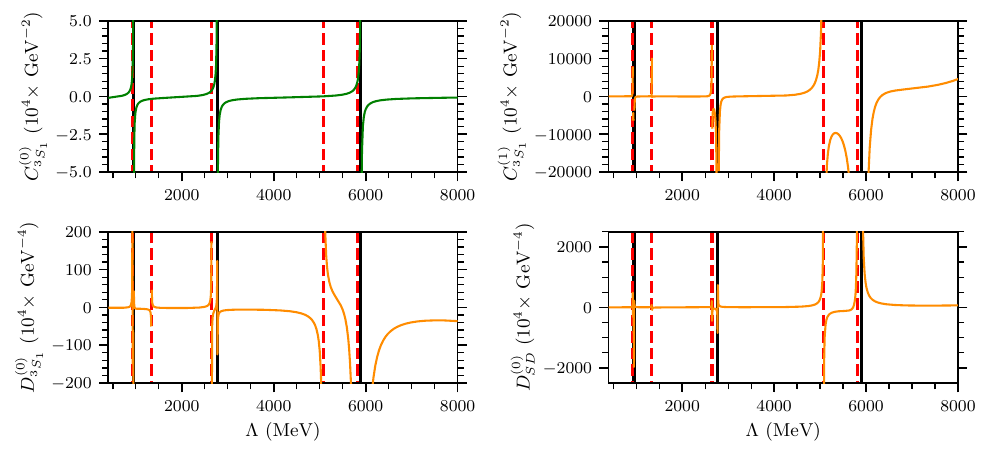}
    \caption{Same as \cref{fig:LECs_3S-D1}, but for a larger cutoff interval.}
    \label{fig:LECs_3S-D1_long}
\end{figure*}

\begin{figure}
    \centering
    \includegraphics[width=0.45\columnwidth]{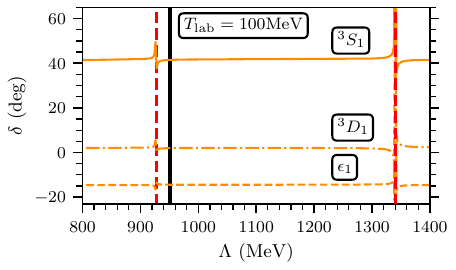}
    \caption{Phase shift predictions in the $\tS$ channel at laboratory kinetic energy $\Tl=100$~MeV as a function of the cutoff, $\Lambda$. The vertical solid line indicates the location of a limit-cycle-like cutoff, while the vertical dashed lines indicate the location of the exceptional cutoffs---in this cutoff interval.}
    \label{fig:3S-D1_phases_Lambda}
\end{figure}

\begin{figure*}
    \centering
    \includegraphics[width=0.95\textwidth]{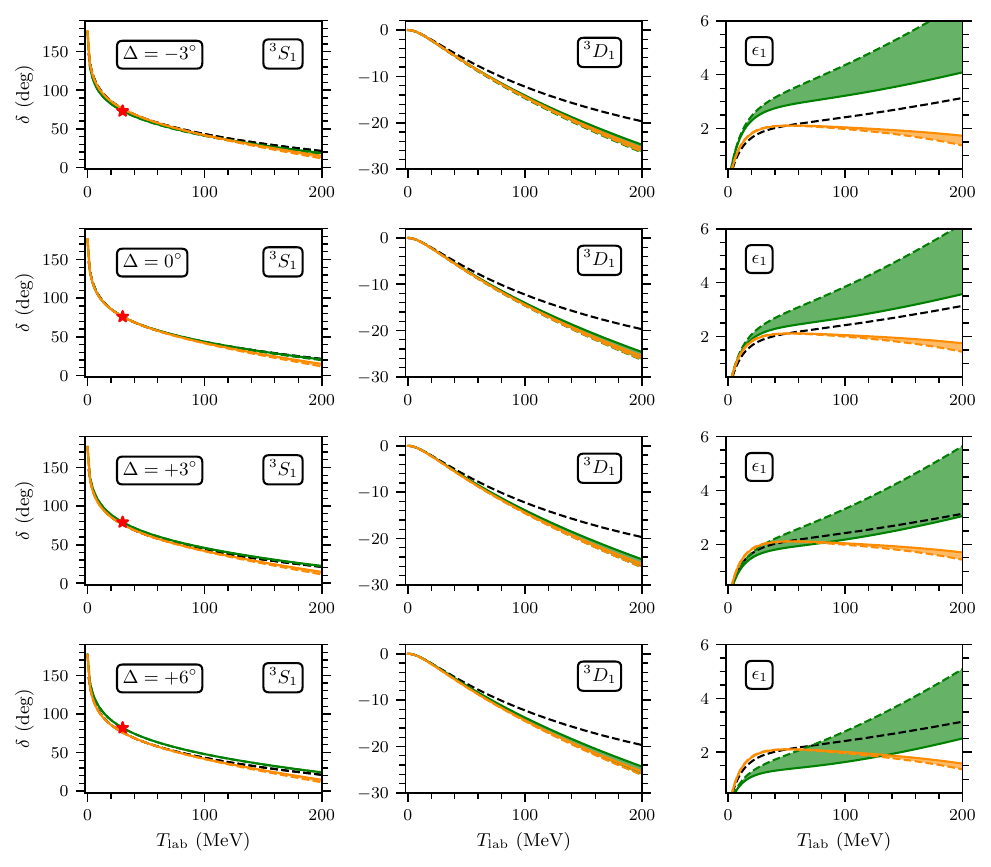}
    \caption{Same as \cref{fig:3S1_shift_phases}, but including an additional shift $(\Delta)$ as well as the $^3D_1$ phase shifts and the mixing angle $\epsilon_1$. The Stapp convention is used \cite{Stapp:1956mz}.}
    \label{fig:3S-D1_shift_phases}
\end{figure*}

\end{widetext}

\end{document}